\documentclass[reqno]{amsart}
\usepackage{graphicx}

\newtheorem{lemma}{Lemma}
\newtheorem{theorem}{Theorem}
\newtheorem{corollary}{Corollary}
\newtheorem{hypothesis}{Hypothesis}
\newtheorem{remark}{Remark}

\newcommand{\be}{\begin{eqnarray}}
\newcommand{\ee}{\end{eqnarray}}
\newcommand{\bee}{\begin{eqnarray*}}
\newcommand{\eee}{\end{eqnarray*}}
\newcommand{\R}{{\mathbb R}}

\newcommand{\N}{{\mathbb N}}
\newcommand{\Z}{{\mathbb Z}}

\newcommand{\I}{\mbox {\sc 1}}
\newcommand{\radius}{l}
\newcommand{\asy}{{\mathcal O}}
\newcommand{\da}{d_A}
\newcommand{\Vector}[1]{\mbox {\bf {#1}}}

\begin{document}

\title {Derivation of the Bose-Hubbard model in the multiscale limit}

\author {Reika FUKUIZUMI}

\address {Graduate School of Information Sciences, Tohoku University, Sendai 980-8579, Japan}

\email {fukuizumi@math.is.tohoku.ac.jp}

\author {Andrea SACCHETTI}

\address {Department of Physical, Information and Mathematical Sciences, University of Modena e Reggio Emilia, Modena, Italy, and Centro S3, Istituto Nanoscienze, CNR, Modena, Italy.}

\email {andrea.sacchetti@unimore.it}

\date {\today}

\thanks {One of us (A.S.) is  grateful for the hospitality of the Tohoku University where part of this paper was written.}

\begin {abstract} In this paper we consider a one-dimensional non-linear Schr\"odinger equation (NLSE) with a periodic potential. \ In the semiclassical limit we prove that the stationary solutions of the Bose-Hubbard equation approximate the stationary solutions of the (NLSE). \ In particular, in the limit of large nonlinearity strength the stationary solutions turn out to be localized on a single lattice site of the periodic potential; as a result the phase transition from superfluid to Mott-insulator phase for Bose-Einstein condensates in a one-dimensional periodic lattice is rigorously proved. 

\bigskip

{\it Ams classification (MSC 2010):} 35Q55; 81Q20 

\bigskip


\end{abstract}

\maketitle

\section {Introduction} \label {Sec1}

Bose-Einstein condensate (BEC) typically consists of a few thousands to millions of atoms which are confined by a trapping potential and at temperature near the absolute zero; and BEC is described by the macroscopic wave function $\psi = \psi (x,t)$ whose time evolution is governed by a self-consistent mean field nonlinear Schr\"odinger equation (NLSE)  \cite {Pi}:
\be
i \hbar \frac {\partial \psi }{\partial t} = H\psi  + \gamma |\psi |^{2\sigma } \psi \, , \ H = - \frac {\hbar^2}{2m} \Delta  + V(x) \, ,\  \label {Equation1}
\ee
with the normalization condition $\int |\psi (\cdot , t )|^2 dx =1$, where $\hbar$ is the Planck constant, $V(x)$ is an external potential, $\gamma \in \R$ is the strength of the nonlinear self-consistent potential and $\sigma >0$ is the nonlinearity power. \ In fact, in the case of cubic nonlinearity, i.e. $\sigma =1$, Eq.(\ref {Equation1}) is usually known as the Gross-Pitaevskii equation (GPE), $\gamma ={\mathcal N} 4 \pi \hbar^2 a_s/m$, where ${\mathcal N}$ is the number of atoms with mass $m$ in the condensate and $a_s$ is the scattering length. 

One of the most important feature of NLSE is the spontaneous symmetry breaking (SSB) effect associated to a bifurcation of stationary solutions. \ Symmetry breaking in ground states of the GPE, with a  symmetric double well linear potential, was considered in \cite {AFGST, FS,GMS,KKSW}. \ In particular, it has been shown that when the total number ${\mathcal N}$ of particles is larger that a critical value ${\mathcal N}_c$ then any ground state is concentrated in only one of the two wells, i.e. the symmetry is broken; the value of ${\mathcal N}_c$ is explicitly given. \ In contrast, for ${\mathcal N}$ smaller than ${\mathcal N}_c$ the ground state is bi-modal, having the symmetries of the linear Schr\"odinger equation. \ A transfer or exchange of stability takes place at such critical value: for ${\mathcal N}<{\mathcal N}_c$ the symmetric state is stable, while for ${\mathcal N}>{\mathcal N}_c$ the symmetric state becomes unstable while the asymmetric state, which raises by bifurcation at ${\mathcal N}={\mathcal N}_c$, is stable. 

This relevant fact opens a new light on the interesting problem of BECs in a lattice and we expect that SSB effects, with the associated transition from delocalized to localized ground states, may occur.  \ A prominent example of such a quantum phase transition is the change from the {\it superfluid phase} to the {\it Mott insulator phase} in a system consisting of bosonic particles in an optical lattice where the external potential is a periodic function \cite {Bl,GMEHB}. \ In fact,  optical lattices basically are arrays of microscopic potentials induced by interfering laser beams and the resulting potential has the form $V(x_1,x_2, \ldots , x_d) = \sum_{\ell =1}^d V_\ell \sin^2 \left (\frac {2\pi}{\lambda_\ell} x_\ell \right )$, where $\lambda_\ell$ is the wavelength of the laser light, corresponding to a lattice periods $a_\ell = \frac 12 \lambda_\ell$, $\ell=1,...d$ \cite {JBCGZ}. \ A first theoretical model describing such a transition has been proposed in the case of cubic nonlinearities (i.e. $\sigma =1$) by the so-called Bose-Hubbard model \cite {FWGF}: 
\be
{\mathbf H} = -  \sum_{k,j \in \Z^d\ : \ |k-j|=1} J_{k,j} {\mathbf b}_j^\dag {\mathbf b}_k + \frac 12 U \sum_k \hat n_k (\hat n_k -1) \label {Equation2}
\ee
where ${\mathbf b}_k$ and ${\mathbf b}_k^\dag$ are the annihilation and creation operators, $[{\mathbf b}_k,{\mathbf b}_j^\dag ]=\delta_j^k$ and $\hat n_k = {\mathbf b}_k^\dag {\mathbf b}_k$ is the number of bosonic atoms at lattice site $k$. \ The boson field operator $\Psi$ is written in terms of the Wannier function $W_1 (x)$ associated to the first band of the periodic Schr\"odinger operator 
\bee
\Psi (x) = \sum_{k \in \Z^d} {\mathbf b}_k W_1 (x-k\cdot a ) \, 
\eee
where $k=(k_1,k_2,\cdots,k_d)$ and $k \cdot a =(k_1 a_1, k_2 a_2, \cdots, k_d a_d)$.
Finally, $U$ can be evaluated as follows
\be
U = \frac {4\pi \hbar^2 a_s}{m} \int_{\R^d}  |W_1 (x)|^4 d x \label {Equation3}
\ee
and
\be
J:= J_{k,j} = - \int_{\R^d} \bar W_1 (x-k\cdot a ) H W_1 (x-j\ \cdot a ) d x \label {Equation4}
\ee
is the hooping matrix element between neighbouring sites $k$ and $j$ such that $|k-j|=1$ (actually it is independent of the indexes). \ Mean field arguments predict that as the strength of the repulsive interaction term $U$ relative to the tunneling term $J$ in the Bose-Hubbard Hamiltonian is changed, the system reaches a critical point in the ratio of $U/J$, for which the system will undergo a quantum phase transition from the superfluid (SF) state to the Mott insulator (MI) state (see also \cite {J} and the references therein, for one-dimensional problem see \cite {Stoferle}). 

Our aim in this paper is to prove the validity of the results predicted by the Bose-Hubbard model in the semiclassical limit for any nonlinearity power $\sigma$ and in dimension $d=1$, and show theoretically the SF-MI transition. \ More precisely, assuming that the potential $V$ is regularly periodic, in the semiclassical limit we prove that stationary solutions of the NLSE (\ref {Equation1}) restricted to the first energy band are given by the stationary solutions of a discrete nonlinear Schr\"odinger equation equivalent to the {\it Bose-Hubbard model} (see Theorem \ref {Theorem0}). 

We would remark that, in the case of cubic nonlinearities (i.e. $\sigma =1$), a mathematical justification of this approximation has been proved in the special case of periodic potentials basically restricted to piecewise constant functions \cite {P}; the extension of such a result to  periodic  potentials like the optical lattice potential $V(x) = V \sin^2 (2 \pi x/\lambda )$ has been given by \cite {PSc} under some technical assumptions on the band functions. \ In this paper, we justify the Bose-Hubbard model for any smooth periodic potentials, without particular assumptions, by making use of the semiclassical analysis for the linear problem  \cite {AH,C}. \ We also prove that the ground state stationary solutions, consisting of orthonormal states on the first Bloch function, are localized on a single well, in the limit of large focusing nonlinearity. \ As a result then the transition from the superfluid phase (for small nonlinearity strength) to the Mott-Insulator phase (for large nonlinearity strength) is proved (see Corollary \ref {Cor1}).

Hereafter, we fix the dimension $d=1$; for the sake of definiteness we assume the units such that $2m=1$; and the semiclassical parameter $\hbar$ is such that $\hbar \ll 1$. \ The notation $\phi \sim \varphi$ means that $\lim_{\hbar \to 0} \frac{\phi}{\varphi}=C$ for some $C\in \R \setminus \{ 0\}$; and $\langle \phi, \varphi\rangle = \int \bar \phi \varphi$; and let us denote $\nabla = \partial_x$ and $\Delta = \partial_{xx}$ even in dimension one; $\nu$ denotes any generic small quantity. \ We will denote eventually a constant $C$ which depends on parameters $\nu, \sigma..$, by $C_{\nu,\sigma..}$ to clarify the dependence on the parameters.

The paper is organized as follows. \ In Section 2 we make use of the semiclassical analysis for the linear problem \cite {AH,C,H,O} in order to construct a semiclassical approximation of the Bloch and Wannier functions with estimate of the remainder terms in the norm of the Banach spaces $L^p$. \ Once we have obtained the semiclassical results in $L^p$ then we apply, in Section 3, a standard argument based on the fixed point method and on the implicit function theorem in Banach spaces \cite {PBook} proving that the stationary solutions of the NLSE can be approximated, in the norm $H^1$, by the stationary solutions of a DNLS, which is the counterpart of the Bose-Hubbard model.

\section {Semiclassical results in dimension 1}

Here, we introduce the assumptions on the periodic potential $V$ and we collect, and extend to the spaces $L^p$, some semiclassical results on the linear operator $H$, formally defined on $L^2 (\R)$ by $H= - \hbar^2 \Delta + V$. We recall that the dimension $d$ is equal to $1$. 

Basically, we assume that 

\begin {hypothesis} \label {Hyp1} $V\in C^2 (\R ) $ is a non-negative regular periodic potential with period $a \in \R$, that is $V(x+ a ) = V(x)$. \ Let 
\bee
L =\left [- \frac 12 a ,\frac 12 a \right )
\eee 
be the lattice. \ We assume that $V(x)$ has a unique non degenerate minimum inside the lattice, i.e.: there exists $x_0 \in L$ such that 
\bee
V(x) > V(x_0) = V_{min}=0 \, , \ \forall x \in L- \{ x_0 \} \, , \ V''(x_0) > 0. 
\eee
\end {hypothesis}
For simplicity, we put $x_0=0$, and we denote by $x_j = j a$ the minima points for $V(x)$ in the following.

\begin {remark}
The assumption $V''(x_0) > 0$ it is only for the sake of definiteness. \ Actually, we could weaken Hyp. \ref {Hyp1} by assuming that $V^{(2m)} (x_0) > 0$ and $V' (x_0) = \ldots = V^{(2m-1)}(x_0)=0$ for some $m \ge 1$. \ In fact, we could admit the existence of two (or more) minima points $x_0$ and $\tilde x_0 \in L$, too, provided that $V^{(m)} (x_0) \not= V^{(m)} (\tilde x_0)$ for some $m \ge 2$.
\end {remark}

It is well known that the spectrum of $H$ is given by bands. \ Let $b=2\pi /a$ be the period of the reciprocal lattice 
\bee
{\mathcal B} =\left [- \frac 12 b ,\frac 12 b \right )  \, . 
\eee
For any fixed value of the quasimomentum variable $\kappa \in {\mathcal B}$ the spectral problem 
\bee
H \varphi = E \varphi 
\eee
with quasi-periodic boundary condition
\bee
\varphi (x+a , \kappa ) = e^{i \kappa a} \varphi (x,\kappa )   
\eee
has purely discrete spectrum with eigenvalues $E_n (\kappa )$; the functions $E_n (k)$ are named \emph {band functions}, and the normalized (on the single lattice cell, with the $L^2$ norm) eigenvectors $\varphi_n (x,\kappa )$ are named \emph {Bloch functions}. \ Both band functions and Bloch functions are periodic functions with respect to $\kappa $, that is $E_n (\kappa +j b)=E_n (\kappa )$ and $\varphi_n (x,\kappa +j b )=\varphi_n (x,\kappa )$ for any $j \in \Z$, and the spectrum of $H$ is given by bands, i.e.  
\bee
\sigma (H ) = \cup_{n \in \N} [\alpha_n , \beta_n ]
\eee
where  
\bee
 [\alpha_n , \beta_n ] = \left \{ E_n (\kappa ) , \ \kappa \in {\mathcal B} \right \} 
\eee
is the $n$-th band; the open intervals $(\beta_n , \alpha_{n+1})$ are named \emph {gaps}.

\begin {remark} \label{remark2} In dimension one, we always have that $\beta_{n} \le \alpha_{n+1} < \beta_{n+1}$ (this is not the case in dimension higher than $1$); and for $\hbar$ small enough it is known that the first bands have its amplitude exponentially small and the gaps between the bands with width of order $\hbar$; that is, for any fixed $N>0$ and any $0< \nu <S_0$ there exist a positive constant $C:=C_{\nu , N}$ independent of $\hbar$, and $\tilde{\hbar}=\tilde{\hbar}_N>0$ such that  
\bee
|\beta_n - \alpha_n | \le C e^{-(S_0-\nu)/\hbar} \ \mbox { and } \ C \hbar \le |\alpha_{n+1} - \beta_n | \le \frac {1}{C} \hbar 
\eee
for any $n=1,2,\ldots , N$ and any $\hbar \in (0,\tilde{\hbar})$; $S_0$ is the Agmon distance between two adjacent sites defined in the equation (\ref {Equation10}) below. \  In particular the first $N$ gaps are all open and the $n$-th band is centered at the $n-$th eigenvalue $\lambda_n$ of the single well operator $H_0$ defined below in (\ref {Equation5}) (see Theorem 4.3 by \cite {O}). 
\end {remark}

Now, following Carlsson \cite {C} let $H_0$ be the Schr\"odinger operator formally defined on $L^2 (\R)$ by  
\be
H_0 = - \hbar^2 \Delta + \tilde V \label {Equation5} 
\ee
where $\tilde V$ is the potential obtained by \emph {filling} all the wells, but one with center $x_0 =0$ (see \cite {C} for details):
\bee
\tilde V(x) = V(x) + \sum_{j\in \Z - \{ 0 \} } \theta_j (x) \, , \ \theta_j (x):= \theta (x-j a )
\eee
where $\theta (x)$ is a smooth and positive real-valued function with compact support on a neighborhood of $x_0=0$ and such that $V (x) + \theta (x) >\frac {1}{4} \rho^2 $, for any $x\in L$,  for some small and positive $\rho $ independent of $\hbar$. \ Both $V$ and $\tilde V \in L^\infty$, then $H$ and $H_0$ are self-adjoint operators with domain $H^2 (\R)$.

Since the bottom of $\tilde V(x)$ is not degenerate, then the eigenvalue problem associated to the single-well trapping potential $\tilde V(x)$ has spectrum $\sigma (H_0)$ with ground state 
\bee
\lambda_1 =   \sqrt {\frac {V'' (x_0)}{2}} \hbar + \asy (\hbar^2 ) \, , 
\eee
since $V_{min}=0$. \ Furthermore,  
\bee
\mbox {dist} \left [ \lambda_1 , \sigma (H_0) \setminus \{ \lambda_1 \} \right ] > 2C \hbar 
\eee
for some $C>0$ independent of $\hbar$. \ The associated normalized eigenvector $\psi_0 (x)$ exponentially decreases (see, e.g. Thm. 3.2 \cite {BS}), hence $\psi_0 \in L^\infty \cap H^2$. \ From the energy identity 
\bee
\hbar^2 \| \nabla \psi_0 \|_{L^2}^2 \le \hbar^2 \| \nabla \psi_0 \|_{L^2}^2 + \langle \psi_0 , \tilde V \psi_0 \rangle = \lambda_1 \| \psi_0 \|_{L^2}^2 
\eee
it follows that
\be
\| \nabla \psi_0 \|_{L^2} \le C \hbar^{-1/2}. \label {Equation6}
\ee

Furthermore, WKB analysis says that $\psi_0$ is localized in a neighborhood of $x_0=0$ and it behaves like
\be
\psi_0(x) = \hbar^{-1/4}  a (x; \hbar ) e^{-\tilde \da (x,x_0)/\hbar} \, ,  \ a (x;\hbar ) = a_0 (x) + \asy (\hbar ) \label {Equation7}
\ee
where $\tilde \da (x,x_0)$ is the Agmon distance between $x$ and the point $x_0$ defined as 
\bee
\tilde \da (x,x_0) = \left | \int_{x_0}^x \sqrt {\tilde V(s) } ds  \right | \, . 
\eee
By a gauge choice $\psi_0 (x)$ is a positive real-valued function, in particular, $a_0 (x)$ is positive and at least of class $C^2(\R)$: in fact $a_0(x)$ behaves like the first Hermite polynomial in a neighborhood of $x_0=0$ and like $[\tilde V(x)]^{-1/4}$ outside of the neighborhood.

In order to have decay estimates in norm we introduce the following quantity instead of the Agmon distance $\tilde \da$:
\bee
\da^\rho (x,x_0) = \left | \int_{x_0}^x \sqrt { \left [ V(s) - \rho^2 \right ]_+ } ds  \right | \, ,
\eee
and we recall that $\psi_0$ satisfies the following estimate (see eq. (5.1) in \cite {C} where we choose $r=1$)
\be
\left \| e^{\da^\rho (x,x_0)/\hbar} \psi_0 (x) \right \|_{L^2} + \hbar \left \| e^{\da^\rho (x,x_0)/\hbar} \nabla \psi_0 (x) \right \|_{L^2} \le C \, . \label {Equation8}
\ee

Now, let $\da (x,y)$ be the Agmon distance between two points $x$ and $y$ defined as 
\bee
\da (x,y) = \left | \int_x^y \sqrt {V(s)} ds \right | \, . 
\eee
It follows that for any $\nu >0$ we can choose $\rho >0$ such that
\be
(1-\nu) \left [ \da (x,x_0) - \nu \right ] \le \da^\rho (x,x_0) \le \da (x,x_0) \, , \ \forall x \in \R .  \label {Equation9}
\ee
Indeed, by construction it follows that 
\be
\da (x,x_0) - \nu \le \da^\rho (x,x_0) \le \da (x,x_0) \, , \ \forall x \in L. 
\ee
If $x \notin L$ let $x= \tilde x + n a $ where $\tilde x \in L$ and $n \in \Z$ (for argument's sake let us assume that $n >0$ and $\tilde x \ge x_0$). \ Then we can observe that 
\bee
\da^\rho (x,x_0) = \da^\rho (\tilde x , x_0) + n \da^\rho (x_0 , x_0 +a) 
\eee 
and 
\bee 
\da (x,x_0) = \da (\tilde x , x_0) + n \da (x_0 , x_0 +a) \, . 
\eee
Then, it follows that 
\bee
\da^\rho (x,x_0) &\ge& \da (\tilde x, x_0) - \nu + n \da (x_0 , x_0+a) - n \nu \\
&=& \da (x,x_0) - n\nu -\nu 
= d_A(x,x_0)-\frac{n d_A(x_0,x_0+a)}{d_A(x_0,x_0+a)} \nu -\nu\, . 
\eee
If we remark that $\da (x,x_0) \ge n \da (x_0 ,x_0 +a)$ then the last inequality takes the form
\bee
\da^\rho (x,x_0) \ge (1-\nu') \left [ \da (x,x_0) - \nu'' \right ] 
\eee
where we set $\nu' = \nu [ \da (x_0 , x_0 +a)]^{-1} = \asy (\nu )$ and $\nu'' = \nu /(1-\nu') = \asy (\nu )$. \ Finally, Inequality (\ref {Equation9}) follows by denoting $\nu \to \nu := \max [\nu' ,\nu'' ]$.

Let 
\be
S_0 = \inf_{i,j \in \Z ,\ i\not= j} \da (x_i , x_{j})  \ \mbox { and } \ S_0^\rho = \inf_{i,j \in \Z ,\ i\not= j} \da^\rho (x_i , x_{j}) \label {Equation10}
\ee
then, by construction of the potential $V(x)$, it turns out that
\be
S_0 =  \da (x_i , x_{j}) \ \mbox { and } \ S_0^\rho =  \da^\rho (x_i , x_{j}) , \ \mbox { if } \ |i-j|=1\, , \label {Equation11} 
\ee
and 
\be 
S_0 < \da (x_i , x_{j}) \ \mbox { and } \ S_0^\rho < \da^\rho (x_i , x_{j}), \ \mbox { if } \ |i-j|>1 \, , \label {Equation12}
\ee
in particular from an argument similar to the derivation of the first inequality of (\ref {Equation9}) it follows that 
\be
S_0 - \nu \le S_0^\rho \le S_0 \, . \label {Equation13}
\ee

\begin {lemma} \label {Lemma1}
Let $\psi_j (x) = \psi_0 (x-x_j)$, $j \in \Z$. \ It follows that

\begin {itemize}

\item [i.] For some positive constant $C$, independent of $\hbar$ and of the index $j$, 
\be
\| \psi_j \|_{L^p}\le C \hbar^{-  \frac {p-2}{4p} } \, , \ \forall p \in [2,+\infty ] \, .  \label {Equation14}
\ee

\item [ii.]  For any fixed $ \nu>0$ 
\be
\| \psi_j \psi_k \|_{L^p} \le C e^{-\left [  | j-k | (S_0 - \nu' ) - \nu'' \right ] /\hbar } \,  , j\not= k \, , \forall p \in [1, +\infty ] \, , \label {Equation15}
\ee
for some positive constants $C:=C_{\nu}$ and $\nu' , \nu'' = \asy (\nu )$, independent of $\hbar$ and of the indexes $j$ and $k$.

\item [iii.] Finally, let $\chi_0(x)$ be a function with compact support on $x_0=0$, i.e. $\mbox {supp } (\chi_0) =\bar \Omega_0 $ for some open and bounded set $\Omega_0$, then for any $ \nu >0$ 
\be
\| \chi_0 \psi_j \|_{L^p} \le C  e^{-\frac {2}{p} \left [  \da (x_j , \Omega_0) - \tilde \nu \right ] /\hbar} \, , \ \forall p \in [2,+\infty ] \, ,  \label {Equation16}
\ee
for some constant $C:=C_{ \nu}$ and $\tilde \nu = \asy (\nu )$, independent of $\hbar$ and of the index $j \not= 0 $.

\end {itemize}

\end {lemma}

{\it Proof.} \ Indeed, we have that $\| \psi_j \|_{L^2} =1$. \ Then, from (\ref{Equation6}) and from Gagliardo-Nirenberg inequality, 
which holds true for any $p \in [2,+\infty]$, we have the wanted estimate 
\bee
\| \psi_j \|_{L^p } \le C \| \nabla \psi_j \|_{L^2}^{\delta}  \| \psi_j \|_{L^2 }^{1- \delta} \le C \hbar^{-\delta /2}\, , \ \delta = \frac {p-2}{2p}\, . 
\eee
In order to prove (\ref {Equation15})  for any $\nu >0$ let us introduce the function 
\bee
w_0 (x) = e^{(1-\nu ) \da^\rho (x,x_0)/\hbar } \psi_0 (x) 
\eee
which satisfies the following inequality  (see Proposition 3.5.3 by \cite {H})
\be
\| \nabla w_0 \|_{L^2 } + \| w_0 \|_{L^2 } \le C e^{\nu/\hbar}
\label {Equation17}
\ee
for some positive constant $C:=C_\nu$. \ In particular, from this inequality and from the Gagliardo-Nirenberg inequality it follows that
\be
\| w_0 \|_{L^p} \le C \| \nabla w_0 \|_{L^2}^{\delta}  \| w_0 \|_{L^2}^{1- \delta} \le C e^{\nu/\hbar}\, , \ \delta = \frac {p-2}{2p}\, .
\label {Equation18}
\ee

Now, let $w_j(x)=w_0 (x-x_j)$, for which  (\ref {Equation17}) holds true. \ We remark that 
\bee
\da^\rho (x,x_j) + \da^\rho (x,x_k) \ge  \da^\rho (x_j,x_k) = | j-k |  S_0^\rho \ge |j-k| (S_0 - \nu ) \, . 
\eee
We have that 
\bee
\| \psi_j \psi_k \|_{L^p} &=& \| w_j w_k e^{-(1-\nu ) \left [ \da^\rho (x,x_j) + \da^\rho (x,x_k) \right ] /\hbar } \|_{L^p} \\ 
& \le & e^{-(1-\nu ) | j - k | (S_0-\nu)/\hbar}  \| w_j w_k \|_{L^p} \\ 
& \le & e^{-(1-\nu ) | j - k | (S_0-\nu)/\hbar}  \| w_j \|_{L^{2p}} \, \| w_k \|_{L^{2p}} \\ 
&\le & C e^{-\left [ (1-\nu ) | j - k | (S_0 - \nu) -2 \nu \right ] /\hbar}
\eee
hence, the inequality (\ref {Equation15}) follows by defining $(1-\nu )(S_0 - \nu ) = S_0 - \nu'$ and $\nu'' = 2 \nu$ where $\nu' , \nu'' = \asy (\nu )$. 

Finally, in order to prove (\ref {Equation16}) we remark that $\| \chi_0\psi_j \|_{L^\infty} \le C \hbar^{-1/4}$ for some $C>0$. \ Furthermore, we have that
\bee
\| \chi_0 \psi_j \|_{L^2} &=& \| \chi_0 e^{-(1-\nu)\da^\rho (x,x_j)/\hbar } w_j \|_{L^2} \\ 
&\le & \| w_j\|_{L^2}  \| \chi_0 e^{-(1-\nu)\da^\rho (x,x_j)/\hbar } \|_{L^\infty } \\ 
&\le & C e^{\nu/\hbar} e^{-(1-\nu)\da^\rho (\Omega_0,x_j)/\hbar } \\ 
&\le & C e^{\nu/\hbar} e^{-(1-\nu)(1-\nu) [\da (\Omega_0,x_j) - \nu ]/\hbar }
\eee
since $\da^\rho (x,x_j) \ge \inf_{x \in \Omega_0} \da^\rho (x,x_j) =: \da^\rho (\Omega_0 , x_j) $, for any $x \in \Omega_0$, and since (\ref {Equation9}). \ From these inequalities and from the  Riesz-Th\"orin interpolation Theorem
\bee
\| f \|_{L^p} \le \| f \|_{L^2}^{2/p} \| f \|_{L^{\infty}}^{(p-2)/p} , \ \forall f \in L^{2} \cap L^{\infty} \, , \ \forall p \in [2,+\infty ] \, .
\eee
then we have that 
\bee
\| \chi_0 \psi_j \|_{L^p} \le C  e^{\frac {2}{p} \left [ -(1-\nu )(1-\nu) [ \da (x_j , \Omega_0) -\nu ] + \nu \right ] /\hbar} \, , \ \forall p \in [2,+\infty ] \, .
\eee
Then the  result follows by defining $(1-\nu)^2 (\da (x_j , \Omega_0) - \nu ) = \da (x_j , \Omega_0) - \tilde \nu$, where $\tilde \nu = \asy (\nu )$. 
\hfill\qed 
\vspace{3mm}

If $\hbar$ is small enough the first band of $H$ is not degenerate (cf. Remark \ref{remark2}), i.e. $\beta_1 < \alpha_2$, and the restriction of $H$ to the spectral subspace associated to the first band can be described by means of an infinite matrix. \ More precisely, let $\Pi$ be the spectral projection of $H$ on the first band, let $F= \Pi \left [ L^2 (\R) \right ] $, let 
\bee
v_j = \Pi  \psi_j  , \ j \in \Z \, , 
\eee
then we have that 

\begin {lemma} \label {Lemma2}
For any $\nu >0$ and $j\in \Z$ there exists a positive constant $C>0$ independent on $\hbar$ and on the index $j$ such that 
\be
\| \psi_j - v_j \|_{L^2} \le C e^{-(S_0-\nu ) /\hbar} \ \mbox { and } \ \| \nabla (\psi_j - v_j) \|_{L^2} \le C e^{-(S_0-\nu ) /\hbar} \, . \label {Equation19}
\ee
Furthermore
\be
\| \psi_j - v_j \|_{L^p} \le C e^{-(S_0-\nu ) /\hbar} \, , \ \forall p \in [2,+\infty ] \, . \label {Equation20}
\ee
\end {lemma}

{\it Proof.} \ The proof simply follows by adapting the arguments by \cite {C}. \ Indeed, the first estimate in equation (\ref {Equation19}) directly follows by  the result in Equation (5.2) by \cite {C}. \ Concerning the second estimate in (\ref {Equation19}) we apply the result in Equation (5.2) by \cite {C} to the gradient and the estimate (4.11) by \cite {C}. \ Then (\ref  {Equation19}) follows. \ Estimate (\ref {Equation20}) is a consequence of (\ref {Equation19}) and the Gagliardo-Nirenberg inequality. \ The proof is so completed. 
\hfill\qed 
\vspace{3mm}

Let
\bee
d_A^1 (j,k) = d_A (x_j , x_k) 
\eee
and 
\bee 
d_A^2 (j,k) = \inf_{m,n\in \Z, \ j \not= m,n \not= k} \left [ d^1_A (j,m) + d^1_A (m,n) + d^1_A (n,k) \right ] 
\eee
where, by (\ref {Equation11}) and (\ref {Equation12}) and by construction of the potential, it turns out that 
\be
d_A^1 (j,k)  = | j-k | S_0 \ \mbox { and } \ d_A^2 (j,k)  \ge \max [ 2 , | j-k |]  S_0 \, . \label {Equation21}
\ee

The set $\{ v_j\}_{j\in \Z}$ form a basis of $F$ such that (see the Main Theorem in \cite {C} adapted to a regular lattice) for any $\nu \in (0, S_0)$ small enough and fixed there exists a positive constant $C:=C_\nu$ \emph {independent of the indexes $j$ and $k$} such that 
\be
| \langle v_j , v_k \rangle - \delta_j^k | \le C e^{- (1-\nu ) d_A^1(j,k)/\hbar}  \, . \label {Equation22}
\ee
Then 
\bee
\left ( \langle v_j , v_k \rangle \right )_{j,k} = \I + A 
\eee
where $A$ is an Hermitian infinite matrix from $\ell^2$ to $\ell^2$ with 
\be
\| A \| \le C e^{-(S_0-\nu )/\hbar } \, . \label {Equation23}
\ee

Let $B = (b_{j,k})_{j,k}$ be the inverse square root of $\I +A$; the elements $b_{j,k}$ of the matrix $B$ satisfy (see Eq. (5.18) by \cite {C})
\be
b_{j,k} &=& \frac 12 \delta_j^k - \frac 12 \langle \psi_j , \psi_k \rangle + \asy \left ( e^{-(1-\nu) \da^2 (j,k)/\hbar }\right ) \nonumber \\
&=& \delta_j^k + \asy \left ( e^{-[(1-\nu') \da (j,k) - \nu'']/\hbar }\right ) \, . \label {formulaCircle}
\ee
The set of functions 
\bee
u_j = \sum_k b_{j,k} v_k 
\eee
form an orthonormal basis of $F$ and the restriction of $H$ to $F$ on such a basis $\{ u_j \}_j$ is associated to the matrix
\be
\delta_j^k \lambda_1 + w_{j,k} + r_{j,k} \label {Equation24}
\ee
where
\bee
w_{j,k} = - \frac 12 \left [ \langle \psi_j , r_k \rangle + \langle r_j, \psi_k \rangle \right ] \, ,  \ r_j (x) = r_0 (x-x_j)\, , \ 
r_0 = \sum_{m \in \Z ,\ m \not= 0} \theta_m \psi_0  \, , 
\eee
and 
\bee
| r_{j,k} | \le C e^{-(1-\nu ) d_A^2 (j,k)} \, .
\eee

\begin {remark} \label {Remark3} Because $\psi_j$ and $\theta_k$ are real valued functions then it follows that 
\bee
w_{j,k}=w_{k,j} \ \mbox { and } \ w_{j,j}=0 \, . 
\eee
\end {remark}

Here, we state some properties of the vectors $u_j$.

Let $T$ be the translation operator $\left ( T f \right ) (x) = f(x+a)$, where $a$ is the period of $V$. \ Then, by construction it follows that $\psi_j =T^{(j)} \psi_0$ and, since $[H , T ]=0$, $[\Pi , T]=0$, hence $v_j =T^{(j)} v_0$.  

\begin {lemma} \label {NewRemark}
Let $T$ be the translation operator $\left ( T f \right ) (x) = f(x+a)$, where $a$ is the period of $V$, then $u_j =T^{(j)} u_0$.
\end {lemma}

{\it Proof.} \ First of all we observe that the matrix $A$ is an Hermitian matrix of Toeplitz type (i.e. $a_{j,k} = a_{j-k,0}$), indeed the elements of the matrix $A$ are such that
\bee
\delta_j^k + a_{j,k} = \langle v_j , v_k \rangle = \langle v_0 , T^{k-j} v_0 \rangle \, , 
\eee
that is, $a_{j,k} = a_{j-k,0}:=a_{j-k}$ for some sequence $\{ a_{\ell } \}_{\ell \in \Z}$ such that $a_{-\ell}= \bar a_\ell$. \ Hence, the matrix $B=[\I + A]^{-1/2}$ is defined as the convergent series (indeed $\|A \| <1$)
\bee
B = \sum_{n=0}^\infty c_n A^n , \ c_n = \frac {(-1)^n (2n)!}{2^{2n} (n!)^2} \sim \frac {1}{\sqrt {n}} \, ; 
\eee
where, by means of a straightforward calculation, it follows that $B(1+A)B = \I$. \ Therefore, the matrix $B$ is of Toeplitz type, since the power $A^n$, $n \in \N$, of the Toeplitz matrix $A$ is still a Toeplitz matrix. \ That is the elements of the matrix $B$ are such that $b_{j,k} = b_{j-k}$ for some sequence $\{ b_{\ell } \}_{\ell \in \Z}$ such that $b_{-\ell}= \bar b_\ell$. \ From these facts it follows that
\bee
u_0 = \sum_k b_{-k} v_k \ \mbox { and } \ u_j = \sum_k b_{j-k} v_k \, ; 
\eee
in particular, it follows that
\bee
T^{(j)} u_0  = \sum_k b_{-k} T^{(j)} v_k  = \sum_k b_{-k} v_{k +j} = \sum_m b_{j-m} v_{m} = u_j \, . 
\eee
\hfill\qed 
\vspace{3mm}

\begin {lemma} \label {Lemma3} For any $\nu>0$ and $j\in \Z$ there exists a positive constant $C>0$ independent on $\hbar$ and on the index $j$ such that 
\be
\| \nabla \psi_j - \nabla u_j \|_{L^2} \le C e^{-(S_0-\nu ) /\hbar} \, 
\label {Equation25bis}
\ee
and 
\be
\| \psi_j - u_j \|_{L^p} \le C e^{-(S_0-\nu ) /\hbar} \, , \ \forall p \in [2,+\infty ] \, .  \label {Equation25}
\ee
\end {lemma}

{\it Proof.} \ Recalling that $\psi_j =T^{(j)} \psi_0$, $v_j = T^{(j)} v_0$ and $u_j =T^{(j)}u_0$, where $T$ is the translation operator defined in Remark \ref {NewRemark}, then we can restrict  ourselves to $j=0$. \ From (\ref {Equation19}) and (\ref {Equation23}) then 
\bee
\| \psi_0 - u_0 \|_{L^2} \le C e^{-(S_0-\nu) /\hbar } \, 
\eee
for some constant $C$ independent. \ The same estimate follows for the gradient, too:
\bee
\| \nabla (\psi_0 - u_0) \|_{L^2} \le  \| \nabla (\psi_0 - v_0) \|_{L^2} + \| \nabla (u_0 - v_0) \|_{L^2}  
\eee
where it has already been seen that 
\bee
\| \nabla (\psi_0 - v_0) \|_{L^2} \le C e^{-(S_0-\nu) /\hbar }.
\eee
The second term is written as 
\bee
\nabla (u_0 - v_0) = \sum_{k} b_{0,k} \nabla v_k - v_0 = \sum_k \left [ b_{0,k} - \delta_0^k \right ] v_k 
\eee
and then 
\bee
\| \nabla (u_0 - v_0) \|^2_{L^2} &\le & \sum_k \left | b_{0,k} - \delta_0^k \right |^2 \| \nabla v_k \|^2  \\ 
&\le & \sum_k \left | b_{0,k} - \delta_0^k \right |^2 \left [ 2 \| \nabla \psi_k \|^2 + 2 \| \nabla (\psi_k - v_k )\|^2 \right ].
\eee
Here the right hand side may be estimated as follows.
\bee
\| \nabla \psi_k \|_{L^2} \le C \hbar^{-1/2}, \quad \| \nabla (v_k - \psi_k )\|_{L^2} \le C e^{-(S_0-\nu) /\hbar }
\eee
and 
\bee
\sum_{k} \left | b_{0,k} - \delta_0^k \right |^2 \le \sum_{k} \left | b_{0,k} - \delta_0^k \right | = \| \I - B \| \le C e^{-(S_0-\nu) /\hbar } \, . 
\eee
These inequalities imply (\ref{Equation25bis}) and from the Gagliardo-Nirenberg inequality then (\ref {Equation25}) follows. \ The proof is so completed.
\hfill\qed 
\vspace{3mm}

\begin {lemma} \label{lem:realvalued} 
The functions $u_j$ and $v_j$ can be chosen to be real-valued.
\end {lemma}

{\it Proof.} \ We start proving that $v_j$ are real-valued, because the functions $\psi_j$ are positive. \ Indeed  $v_j = \Pi \psi_j$ where $\Pi$ is the projection operator on the first band.\ From the Bloch decomposition formula (see eq. (2.1.22) by \cite {PBook}) it follows that 
any vector $\psi \in L^2$ can be written as 
\bee
\psi (x) = \sum_{n \in \N} \int_{{\mathcal B}} \varphi_n (x,\kappa ) a_n (\kappa ) d \kappa  
\eee
where ${\mathcal B}$ is the Brillouin zone, $\kappa $ is the quasimomentum (or also crystal momentum) variable and $\varphi_n (x,\kappa )$ are the Bloch functions. \ The function $a_n (\kappa )$ is called the crystal momentum representation of the wave vector associated 
to $\psi$  and it is defined as
\bee
a_n (\kappa ) = \int_{\R} \bar \varphi_n (x, \kappa ) \psi (x)  d x 
\eee
Then, the restriction to the first band is simply given by for any vector $\psi \in L^2$
\bee
\left (\Pi \psi  \right ) (x) = \int_{{\mathcal B}} \varphi_1 (x,\kappa ) a_1 (\kappa ) d \kappa  
\eee
and it is real valued; indeed, 
\bee
\overline {\left (\Pi \psi  \right ) (x) } &=& \int_{{\mathcal B}} \bar \varphi_1 (x,\kappa ) \overline {a_1 (\kappa )} d \kappa \\ 
&=&  \int_{{\mathcal B}} \varphi_1 (x,-\kappa ) \overline {a_1 (\kappa )} d \kappa \\ 
&=&  \int_{{\mathcal B}} \varphi_1 (x,\kappa ) \overline {a_1 (-\kappa )} d \kappa 
\eee
since (see eq. (2.1.17) by \cite {PBook})) $\varphi_n (x,-\kappa ) = \bar \varphi_n (x,\kappa )$. Moreover, if $\psi (x)$ is a real-valued function,
\bee
\overline {a_1 (-\kappa )} &=&  \int_{\R} \varphi_1 (x, -\kappa ) \bar \psi (x)  d x = \int_{\R} \bar \varphi_1 (x, \kappa ) \bar{\psi} (x)  d x 
= \int_{\R} \bar \varphi_1 (x, \kappa ) \psi (x)  d x 
= a_1 (\kappa ).  
\eee
We can apply this argument for $\psi=\psi_j$. \ Once we have proved that $v_j$ are real-valued functions then immediately follow that the elements of the matrices $A$ and $B$ are real-valued, and then, by construction, $u_j$ are real valued functions too.
\hfill\qed 
\vspace{3mm}

\begin {lemma} \label {lemma10} The wavefunctions $u_j$ are such that:

\begin {itemize} 

\item [i.] $ \| u_j u_k \|_{L^1} \le C e^{-[(S_0-\nu' )|j-k|- \nu'']/\hbar } $, $j\not= k$, for any $\nu' , \ \nu'' >0$ and for some positive constant $C>0$ independent on the indexes $j$ and $k$;

\item [ii.] $\| \sum_j |u_j| \|_{L^\infty} \le C \hbar^{-1/2}$ for some $C>0$.

\end {itemize}

\end {lemma}

{\it Proof.} \ In order to prove property i. we recall the following results by \cite {C} (see, respectively, the proofs of Lemmata 5.1 and 5.2):
\bee
\left \| e^{\left [ \da^\rho (x_j, \cdot) + \da^\rho (x_k, \cdot ) \right ]/\hbar } \psi_j (\cdot ) \psi_k (\cdot ) \right \|_{L^1} \le C 
\eee
and
\bee
\left \| e^{\left [ \tilde \da^\rho (x_j, \cdot) + \tilde \da^\rho (x_k, \cdot ) \right ] /\hbar} \left [ \psi_j (\cdot ) - v_j (\cdot ) \right ] \, \left [ \psi_k (\cdot ) - v_k (\cdot ) \right ] \right \|_{L^1} \le C 
\eee
and
\be
\left \| e^{\tilde \da^\rho (x_j, \cdot)/\hbar} \left [ \psi_j (\cdot ) - v_j (\cdot ) \right ]  \right \|_{L^2} \le C \label {formulaStar} 
\ee
where 
\bee
\tilde \da^\rho (x_j , x) = \inf_{\ell \not= j} \left [ \da^\rho (x_j , x_\ell) + \da^\rho (x_\ell , x) \right ] \ge \da^\rho (x_j ,x) \, . 
\eee
and $C$ is a positive constant independent of the indexes $j$ and $k$. \ Hence, it follows that 
\bee
&& \left \| e^{\left [ \tilde \da^\rho (x_j, \cdot) + \tilde \da^\rho (x_k, \cdot ) \right ]/\hbar } v_j (\cdot ) v_k (\cdot ) \right \|_{L^1}  \\		
&& \ \ \le \left \| e^{\left [ \tilde \da^\rho (x_j, \cdot) + \tilde \da^\rho (x_k, \cdot ) \right ]/\hbar } \psi_j (\cdot ) \psi_k (\cdot ) \right \|_{L^1}  \\  
&& \ \ + \left \| e^{\left [ \tilde \da^\rho (x_j, \cdot) + \tilde \da^\rho (x_k, \cdot ) \right ]/\hbar } [\psi_j (\cdot )-v_j(\cdot )] \psi_k (\cdot ) \right \|_{L^1}  \\ 
&& \ \ + \left \| e^{\left [ \tilde \da^\rho (x_j, \cdot) + \tilde \da^\rho (x_k, \cdot ) \right ]/\hbar } [\psi_k (\cdot )-v_k(\cdot )] \psi_j (\cdot ) \right \|_{L^1}  \\ 
&& \ \ + \left \| e^{\left [ \tilde \da^\rho (x_j, \cdot) + \tilde \da^\rho (x_k, \cdot ) \right ]/\hbar } [\psi_j (\cdot )-v_j(\cdot )]\, [\psi_k (\cdot )-v_k(\cdot )] \right \|_{L^1} \le C 
\eee
from which it follows that $\| v_j v_k \|_{L^1} \le C e^{-[(S_0-\nu' )|j-k|- \nu'']/\hbar } $ for any $\nu' , \ \nu'' >0$ and for some $C>0$. \ In order to get the estimate on $u_j$ we recall the estimate (\ref {formulaCircle}), from this fact and by means of a straightforward calculation then estimate i. follows.

In order to prove property ii. we remark that the sum $\sum_j |u_j (x)|$, if convergent, defines a periodic function with period $a$ since $u_j (x) = u_0(x+aj)$; hence, we may restrict the $L^\infty $ estimate of such a sum on the single interval with length $a$. \ Now, let $n$ be any integer number and we set $I_n =[(n-1/2) a, (n+1/2)a]$, where $a$ is the period of the potential $V(x)$, and we are going to estimate $u_0 (x)$ in the interval $I_n$. \ Then we observe that 
\bee
\| u_0 \|_{L^\infty (I_n)} \le \sum_m |b_{0,m} \| v_m \|_{L^\infty (I_n)} 
\eee
where the coefficients $b_{0,m}$ of the matrix $B$ satisfy (\ref {formulaCircle}) and where $\| v_m \|_{L^\infty (I_n)} $ can be estimated by means of the Gagliardo-Nirenberg inequality:
\bee
\| v_m \|_{L^\infty (I_n)} \le C \left [ \| \nabla v_m \|_{L^2 (I_n)} + \| v_m \|_{L^2 (I_n)} \right ] 
\eee
where 
\bee
 \| \nabla v_m \|_{L^2 (I_n)} \le \| \nabla v_m \|_{L^2 (\R )} \le C \hbar^{-1/2}
\eee
and where $\| v_m \|_{L^2 (I_n)}$ can be estimated as follows: 
\bee
\| v_m \|_{L^2 (I_n)} &=& \left \| e^{-g_m (x)} \left [ e^{g_m(x)} v_m \right ] \right \|_{L^2(I_n)} \\ 
&\le & \| e^{-g_m (x) } \|_{L^\infty (I_n)} \left \|  e^{g_m(x)} v_m  \right \|_{L^2(I_n)} \\
& \le & C e^{-(S_0-\nu ) |n-m|/\hbar } 
\eee
where $g_m(x) = \tilde \da^\rho (x_m,x)/\hbar $ and where 
\bee
\left \|  e^{g_m (x)} v_m  \right \|_{L^2(I_n)} \le \left \|  e^{g_m (x)} v_m  \right \|_{L^2(\R )} \le C
\eee
since (\ref {Equation8}) and (\ref {formulaStar}). \ Collecting all these results then it follows that 
\bee
 \| u_j \|_{L^\infty (I_n)} \le C e^{-(S_0 - \nu) |n-j|/\hbar } \, , \ j \not= n , 
\eee
from which property ii. follows:
\bee
\left \| \sum_j |u_j| \right \|_{L^\infty (\R )} 
&=&  \left \| \sum_j |u_j| \right \|_{L^\infty (I_0 )} \le  \| u_0 \|_{L^\infty (I_0 )} +  \| \sum_{j\not= 0} |u_{j}| \|_{L^\infty (I_0 )} \\ 
&\le & \|  u_0 \|_{L^\infty (\R )} + \sum_{j\not= 0} C e^{-(S_0 - \nu) |j|/\hbar } \le C \hbar^{-1/2} + C  e^{-(S_0 - \nu) /\hbar } \, . 
\eee
\hfill\qed 
\vspace{3mm}

\begin {remark} \label {Remark5}
The wave-vectors $u_j (x)$ construct the first Bloch function $\varphi_1 (x,\kappa)$. \ More precisely, let $\chi_0 (x)$ be a function with compact support contained in an open set $M \subset B_{x_0} (S_0)$ such that $0 \le \chi_0 \le 1$ and it is exactly $1$ on $B_{x_0} (S_0-\alpha )$ for some $\alpha >0$ fixed. \ Let $\chi_j (x) = \chi_0 (x-x_j)$. \ Then, the vector 
\bee
\phi_1 (x,\kappa) = \sum_{j\in \Z} e^{i \kappa \cdot j a} \chi_j (x) u_j (x), \quad \kappa \in \mathcal{B} 
\eee
well approximates the Bloch function $\varphi_1 (x,\kappa)$ in the sense of Lemma 3.3 by \cite {O}. \ Hence, the Wannier function $W_1(x)$ defined as 
\bee
W_1 (x) = \frac {1}{|{\mathcal B}|} \int_{\mathcal B} \varphi_1 (x,\kappa) d \kappa 
\eee
is well approximated by 
\bee
\frac {1}{|{\mathcal B}|} \int_{\mathcal B} \phi_1 (x,\kappa) d \kappa = \chi_0 (x) u_0(x)  = \psi_0 (x)  + \asy (e^{-(S_0-\alpha ) /\hbar } )  
\eee
because of Lemma \ref {Lemma3}.
\end {remark}

Furthermore, the following results concerning the elements $w_{j,k}$ hold true.

\begin {lemma} \label {Lemma4} Let 
\bee
\beta = - w_{j,k} \ \mbox { if } \ |j-k|=1
\eee
then $\beta >0$ is independent of the indexes $j$ and $k$ and for any $\nu >0$ then 
\be
\frac 1C e^{-(S_0+\nu )/\hbar} \le \beta \le C e^{-(S_0-\nu )/\hbar}  \label {Equation26}
\ee
for some positive constant $C:=C_\nu >0$.
\end {lemma}

{\it Proof.} \ By the symmetry of the periodic potential then immediately follows that $w_{i,j}$ is independent of the indexes when $|i-j|=1$, then we restrict ourselves to the cases of $j=0$ and $k=1$. \ Moreover, since $\psi_0$ and $\theta_1$ are positive functions then $\beta $ is a positive quantity. \ In order to get the inequality (\ref {Equation26}) we have to estimate the quantity
\bee
\langle \psi_0 , r_1 \rangle &=& \int_{\R} \psi_0 (x) r_1 (x) dx  \\ 
&=& \sum_{m \not= 1}  \int_{\R} \psi_0 (x) \theta_m (x) \psi_1 (x) dx \\ 
&=& \int_{\R} \psi_0 (x) \theta_0 (x) \psi_0 (x-x_1) dx + \sum_{m\not= 0,1} \int_{\R} \psi_0 (x) \theta_m (x) \psi_1 (x) dx 
\eee
Now, let $\chi_0 = \sqrt {\theta_0}$ where the support of $\chi_0$ is contained in an open set $\Omega_0$ such that $\da (x_j, \Omega_0) \ge \frac {2}{3} S_0$ for any $j\not= 0$, then by (\ref {Equation16}) it follows that 
\bee
\sum_{m\not= 0,1} \int_{\R} \psi_0 (x) \theta_m (x) \psi_1 (x) dx &=& \sum_{m\not= 0,1} \int_{\R} \psi_{-m} (x) \theta_0 (x) \psi_{1-m} (x) dx \\ 
&= & \sum_{m\not= 0,1} \int_{\R} \chi_0 (x) \psi_{-m} (x) \chi_0 (x) \psi_{1-m} (x) dx \\
&\le & \sum_{m\not= 0,1} \| \chi_0 (\cdot ) \psi_{-m} (\cdot ) \|_{L^2} \, \| \chi_0 (\cdot ) \psi_{1-m} (\cdot ) \|_{L^2} \\
&\le & \sum_{m\not= 0,1} C e^{2\tilde \nu /\hbar} e^{- \left [ \da (x_{-m},\Omega_0) + \da (x_{1-m}, \Omega_0) \right ] /\hbar } \\ 
&\le & C e^{- \left [ \frac 43 S_0 - 2 \tilde \nu \right ]/\hbar } 
\eee
for some $\tilde \nu <\frac 13 S_0$ provided $\nu$ is small enough (here $\tilde \nu = O(\nu) $) . \ Finally, from (\ref {Equation15}) it follows that 
\bee
\int_{\R} \psi_0 (x) \psi_1 (x) \theta_0 (x) d x \le  C \| \psi_0 \psi_1 \|_{L^1} \le C e^{-\left [  (S_0 -\nu )- \tilde \nu \right ]/\hbar } \, , 
\eee
hence the r.h.s. of (\ref {Equation26}) follows. \ In order to prove the l.h.s. of (\ref {Equation26}) let  
\bee
B_{x_0} (\radius ) = \{ x \in \R \ : \ \da (x,x_0) \le \radius \}
\eee
be the interval with center $x_0=0$ and radius $\radius >0$, and let $\radius_1>0$ be such that $B_{x_0} (\radius_1) \subseteq \mbox {supp} (\theta_0)$ and 
\bee
\int_{\R} \psi_0 (x) \psi_1 (x) \theta_0 (x) d x \ge C \int_{B_{x_0} (\radius_1)} \psi_0 (x) \psi_0 (x-x_1) d x
\eee 
In order to estimate the integral on $B_0 (\radius_1)$, let $\alpha >0$ be fixed and let 
\bee
\Gamma_\alpha  = \{ x \ : \ \da^\rho (x,x_0)+\da^\rho (x,x_1) \le S_0+ \alpha \} \, , 
\eee
where we underline that this set is not empty provided $\rho$, which enters in the definition of the function $\theta (x)$, is small enough. \ Then, from the WKB expansion of $w_0 (x; \hbar ) = \hbar^{-1/4} \left [ a_0 (x) + \asy (\hbar ) \right ] $ on the set $[-2L,2L]$ we have that  
\bee
&& \int_{B_{x_0} (\radius_1)} \psi_0 (x) \psi_0 (x-x_1) d x= \\ 
&& \ =  \int_{B_{x_0}(\radius_1)} w_0 (x;\hbar ) w_0 (x-x_1;\hbar ) e^{-(\da^\rho (x,x_0)+\da^\rho (x,x_1))/\hbar} dx \\ 
&& \ =  \int_{B_{x_0}(\radius_1) \cap \Gamma_\alpha } w_0 (x;\hbar ) w_0 (x-x_1;\hbar ) e^{-(\da^\rho (x,x_0)+\da^\rho (x,x_1))/\hbar} dx + \\ 
&& \ \ + \int_{B_{x_0}(\radius_1) \setminus \Gamma_\alpha } w_0  (x;\hbar ) w_0 (x-x_1;\hbar ) e^{-(\da^\rho (x,x_0)+\da^\rho(x,x_1))/\hbar} dx \\ 
&& \ \ge  \int_{B_0(\radius_1) \cap \Gamma_\alpha } w_0 (x;\hbar ) w_0 (x-x_1;\hbar ) e^{-(\da^\rho(x,x_0)+\da^\rho (x,x_1))/\hbar} dx \\ 
&& \ \ge 2 c  | B_{x_0}(\radius_1) \cap \Gamma_\alpha | \hbar^{-1/2} e^{ -(S_0+\alpha)/\hbar}
\eee
where the measure of the set $ B_{x_0}(\radius_1) \cap \Gamma_\alpha$ is not zero and where
\bee
c := \min_{x \in B_{x_0}(\radius_1)} a_0 (x) a_0 (x-x_1) >0 \, .
\eee
The Lemma is proved.
\hfill\qed 
\vspace{3mm}

\begin {lemma} \label {Lemma5} If $|j-k|>1$ then for any $\nu >0$ there exists a positive constant $C:=C_\nu$ independent of the indexes $j$ and $k$ such that 
\bee
|w_{j,k}| \le C e^{- \left [ | j-k | (S_0-\nu ) - \frac 23 S_0 \right ] /\hbar }  \, .  
\eee
\end {lemma}

{\it Proof.} \ The proof would make use of the same arguments of Lemma \ref {Lemma4}. \ In order to estimate the term $\langle \psi_j , r_k \rangle $ we observe that is equal to the term $\langle \psi_{j-k} , r_0 \rangle $ and thus we can restrict ourselves to terms of the type $\langle \psi_k , r_0 \rangle $, where $|k|>1$. \ We have that 
\bee
|\langle \psi_k , \theta_m \psi_0 \rangle | &=& |\langle \psi_{k-m} , \theta_0 \psi_{-m} \rangle | \le  \| \chi_0 \psi_{k-m} \|_{L^2} \, \| \chi_0 \psi_{-m} \|_{L^2} \\ 
&\le & C e^{2\tilde \nu /\hbar} e^{- \left [ \da (x_{k-m}, \Omega_0 ) + \da (x_{-m}, \Omega_0 ) \right ] /\hbar } 
\eee
where $\tilde \nu = O(\nu )$ and 
\bee
\da (x_{-m} , \Omega_0 ) \ge \left ( | m | - \frac 13 \right ) S_0 \ \mbox { and } \ \da (x_{k-m} , \Omega_0 ) \ge \left ( | k-m | - \frac 13 \right ) S_0 \, . 
\eee
Thus, we have to consider the sum 
\bee
\sum_{m\not= 0} e^{- \left [ | m | + | k-m | - \frac 23 \right ] S_0 /\hbar }\,, \ | k | >1, 
\eee
which can be estimated by $Ce^{-\left [ |k| (S_0 - \nu ) - \frac 23 S_0 \right ] /\hbar}$. \ The Lemma is proved.
\hfill\qed 
\vspace{3mm}

\begin {remark} \label {Remark4}
The matrix obtained collecting the elements $(w_{j,k})$, for $j,k\in \Z$ such that $|j-k|>1$, and the elements $r_{j,k}$, for $j,k \in \Z$, defines a linear operator 
\bee
\tilde D : \ell^p (\Z) \to \ell^p (\Z) \, . 
\eee
This operator is bounded for any $p \in [1,+\infty ]$ with bound 
\bee
\| \tilde D \|_{{\mathcal L} (\ell^p  \to \ell^p )} \le C e^{-(S_0 + \alpha )/\hbar }\, ,
\eee
for some positive constant $0<\alpha <S_0$ independent of $\hbar$ and $p$, and for some positive constant $C$ dependent on $p$ and independent of $\hbar$.
\end {remark}

\section {Derivation of the Bose-Hubbard model}

Here, we consider the nonlinear Schr\"odinger equations
\be
i \hbar {\partial_t \psi } = H \psi + \gamma |\psi |^{2\sigma}  \psi 
\label {Equation27}
\ee
where $\gamma \in \R$, 
\be
H = - {\hbar^2} \Delta + V  \, ,  \label {Equation28}
\ee
is the linear Hamiltonian with a \emph {periodic} potential $V(x)$, and $|\psi |^{2\sigma} \psi $ is a nonlinear perturbation; $\gamma$ denotes the strength of the nonlinear perturbation and it is a real number, here we consider both attractive/focusing (when $\gamma<0$) and repulsive/defocusing (when $\gamma>0$) cases.

We introduce the \emph {effective nonlinearity parameter} defined as
\bee
\eta := \frac {C_0 \gamma}{\beta}
\eee
where $C_0=\| u_0\|_{L^{2\sigma+2}}^{2\sigma+2}$. \ We remind that the set $\{u_j\}_{j\in \Z}$ form an orthonormal  
basis of the space projected on the first band.

\begin {remark} {We point out that, in the case of cubic nonlinearity corresponding to $\sigma =1$, the effective nonlinearity parameter $\eta$ coincides, up to corrections of order $\hbar$, with the ratio $U/J$ between the two parameters defined in the Bose-Hubbard model (\ref {Equation2}). \ Indeed, in the case of $\sigma =1$ we have that 
\bee
C_0 = \| u_0 \|_{L^{2\sigma +2}}^{2\sigma +2} \sim \| \psi_0 \|_{L^{2\sigma +2}}^{2\sigma +2} = \int |\psi_0 (x)|^4 dx \sim \int |W_1 (x)|^4 dx
\eee
since Remark \ref {Remark5}. \ Hence, the leading term of $\gamma C_0$ coincides with the value of the parameter $U$ defined by  (\ref {Equation3}). \ Furthermore, the dominant term of the hopping matrix element $J$ defined by (\ref {Equation4}) coincides with $\beta$
\bee
J &=& - \int \bar W_1 (x-x_1) H W_1 (x) dx = - \int \bar W_1 (x-x_1) \left [ - {\hbar^2} \Delta + V \right ]  W_1 (x) dx \\ 
& \sim & - \int \bar \psi_0 (x-x_1) \left [ - {\hbar^2} \Delta + V \right ]  \psi_0 (x) dx \\ 
& \sim & - \int \bar \psi_0 (x-x_1) \left [ \sum_{j\not= 0} \theta_j (x) \right ]  \psi_0 (x) dx  = \beta 
\eee
in the limit of $\hbar$ small enough because of (\ref {Equation7}).}
\end {remark}

\begin {hypothesis} \label {Hyp2} [Multi-scale limit] We consider the simultaneous limit of small $\hbar$ and small $\gamma$ such that the \emph {effective nonlinearity parameter} $\eta$  goes to a finite value.
\end {hypothesis}

\begin {remark} \label {Remark7} Since $\| \psi_0 \|_{L^{2\sigma +2}}^{2\sigma +2} \le C \hbar^{-\sigma /2}$ and the hopping matrix parameter between neighbouring wells $\beta$ is exponentially small as $\hbar$ goes to zero, then Hypothesis \ref {Hyp2} implies that $\gamma$ exponentially goes to zero as $\hbar$ goes to zero.
\end {remark}

Since $V\in L^\infty$ then the Cauchy problem (\ref {Equation27}) is locally well-posed \cite {CW} and the conservation of the norm of $\psi (x,t)$ and of the energy 
\bee
{\mathcal E} (\psi ) = \langle \psi , H \psi \rangle + \frac {\gamma}{\sigma +1} \|\psi\|_{L^{2\sigma+2}}^{2\sigma+2} \, 
\eee
holds true. \ Furthermore, since $\gamma$ is small enough blow-up cannot occur and, in fact, the Cauchy problem (\ref {Equation27}) is globally well-posed if $|\gamma |$ is small enough. \ Finally, we have a priori estimate 
\be
\| \nabla \psi ( \cdot , t) \|_{L^2} \le C \hbar^{-1/2} \ \mbox { and } \ \| \psi (\cdot , t ) \|_{L^p} \le C \hbar^{-\frac {p-2}{4p}},\quad \forall t\ge 0, \quad \forall p\in [2,\infty] \label {Equation29}
\ee
for some positive constants $C$ (see Theorem \ref {Theorem1} and the associated Remark in \cite{S}).

We set $\psi (x, t) = e^{-i \lambda t /\hbar } \phi (x)$, and we are interested in the associated stationary problem:  
\be
\lambda \phi =  H \phi + \gamma |\phi |^{2\sigma} \phi \, . \label {Equation31} 
\ee

Now, we are going to show that there exists a solution of this stationary problem $\phi\in H^1(\R)$, which can be approximated by means of a Bose-Hubbard model, that is by means of a discrete nonlinear Schr\"odinger equation (DNLS) of the form
\be
E F_k - \sum_{j\in \Z \ : \ |j-k|=1} F_j 
+ \eta |F_k|^{2\sigma} F_k =0 \, , \quad  \| \Vector {F} \|_{\ell^2} =1 \, ,  \label {Equation32}
\ee
where $\Vector {F}=(F_j)_{j\in \Z} \in \ell^2 (\Z )$ and 
\be
E= \frac {\lambda_1 - \lambda}{\beta} \, . \label {Equation33}
\ee

\begin{remark} \label{Remark:existence} The existence of a solution of (\ref{Equation31}) has been investigated by many authors (see \cite{PBook} and references therein). \ For example, the existence of a non trivial solution for the case where $\gamma<0$ and $\lambda< \inf \sigma(H)$ ({\it in semi-infinite gap}) is proved in \cite{PBook} and the case where $\gamma \in \R$, $\lambda > \inf \sigma(H)$ and $\lambda \notin \sigma(H)$ ({\it in a finite gap}) was treated in \cite{Pankov2}. \  No nontrivial solution exists for $\gamma>0$ and $\lambda < \inf \sigma(H)$. \ The approximation that we will show ensures also the existence of a solution $\phi \in H^1(\R)$ of (\ref{Equation31}). \ We remark that by the standard bootstrap argument the solution $\phi$ can gain the regularity, e.g., $\phi\in H^2(\R)$. \ Thus, $\phi$ vanishes at infinity and is also real-valued modulo the gauge transformation  in case of $d=1$ (see Lemma 3.7 of \cite{PBook}). \ Since we consider the problem in one dimension, it suffices to consider a real-valued solution.
\end{remark}

\begin {remark} \label {Remark7Bis} As for (\ref{Equation32}), Weinstein proved in \cite {W} that in the case of attractive nonlinearities, i.e. for $\eta <0$, a minimizer for the associated $\ell^2$-constraint variational problem, in any dimension $d$, 
\bee
I_{\eta}= \inf \Big\{-\langle \delta^2 \Vector{F} , \Vector{F} \rangle_{\ell^2 ({\Z}^d)} -\frac{1}{\sigma+1}\sum_{j} |F_j|^{2\sigma+2} :~ \sum_{j} |F_j|^2 =|\eta|^{1/\sigma} \Big\} 
\eee
exists provided that $\sigma < \frac {2}{d}$, where 
\bee
(\delta^2 \Vector{F})_k = \sum_{j\in \Z^d \ : \ |j-k|=1} F_j - 2 d F_k \, . 
\eee
If $\sigma \ge \frac {2}{d}$ and $\eta<0$, a minimizer exists only when $|\eta|$ is larger than a threshold value $\eta_{thresh} >0$. \ In other words, $\eta_{thresh}$ is a value which ensures $-\infty<I_{\eta}<0$ for all $\eta<0$ with $|\eta| \ge \eta_{thresh}$. \ This threshold value is given by the best constant in the discrete Gagliardo-Nirenberg inequality, 
\bee
\eta_{thresh} = (\sigma +1) {\mathcal I} \, , \quad \ {\mathcal I} = \inf \frac {\left [ \sum_k |F_k|^2 \right ]^\sigma \cdot  \langle - \delta^2 \Vector{F} , \Vector{F} \rangle_{\ell^2 (\Z^d)}} {\sum_k |F_k|^{2\sigma +2} } \, . 
\eee
This result implies the existence of a family of ground states of (\ref {Equation32}) for some $E$. \ It is known that for $E<-2d$ there is no nontrivial solution.
\end {remark}

\begin {remark} \label {Remark7Ter} In the case of $d=1$ the solution $\Vector {F}=(F_j)_{j\in \Z} \in \ell^2(\Z)$ of equation (\ref {Equation32}) can be assumed to be real-valued (see, e.g. Lemma 3.11 by \cite {PBook}); and furthermore it decays exponentially at infinity: i.e. there exist $C>0$ and $\tau>0$ such that  
\bee
|F_j| \le C e^{-\tau |j|}, \quad j\in \Z, 
\eee
(see Theorem 1.1 in \cite {Pankov}). \ In particular $\Vector {F} \in \ell_1^2 (\Z) \subset \ell^1(\Z)$, where 
\bee
\ell_1^2 (\Z) = \left \{ \Vector{c}=(c_j)_{j\in \Z},~ \sum_{j\in \Z} (1+|j|^2)|c_j|^2 <\infty \right \}.
\eee
Also, the case $d=1$ allows us to have much more information than the case of $d\ge 2$. \ The existence of two families of positive localized modes, known as the site-symmetric solution, the bond-symmetric solution, is established in \cite{QX} for the case $E>2$, with the use of a method in the dynamical system. \ Analysis of the anticontinuum limit has been done by a simple application of Implicit Function Theorem in \cite[Theorem 3.8]{PBook}; for $|\eta |$ large enough then (\ref {Equation32}) admits a family of ground state with energy $E=2 - \eta + o(1)$ in the limit of large $|\eta |$ and with wavevector $(F_j) \sim \delta_j^{j_0}$ for any $j_0 \in \Z$.
\end{remark}

We will need the following assumption too. Recall that we fix the dimension $d=1$.  

\begin {hypothesis} \label {Hyp3} 
Let $\Vector{F}=(F_j)_{j\in \Z}$ be a real-valued solution of DNLS (\ref {Equation32}). \ Then, we assume that the linearized map at $\Vector{F}$
\bee
L_+ : \ell^1_\R (\Z ) \to \ell^1_\R (\Z) , \ \ \ell^1_{\R} := \{ \Vector {c} = \{ c_j \} \in \ell^1(\Z) \ :\ c_j \in \R \}\, , 
\eee
defined as 
\be
(L_+ \Vector{c})_j &=& -(c_{j+1}+c_{j-1} ) +(E + \eta (2\sigma+1) |F_j|^{2\sigma}) c_j,\, \ \Vector{c} \in \ell^1_\R (\Z) \, ,    \label {Equation33a}
\ee
is one-to-one and onto.
\end {hypothesis}

The main theorem in this section is the following. 

\begin{theorem} \label{Theorem0}
Let Hypotheses 1-3 be satisfied. Assume $\hbar >0$ small enough and $\sigma \ge 1/2$. \ Let $E$ be such that (\ref{Equation32}) admits a solution, and fix $\Vector{F} \in \ell^1(\Z)$ one real-valued solution of DNLS (\ref{Equation32}) 
associated to $E$. \ Then, there exists $C_{\hbar}>0$ such that Eq. (\ref {Equation31}) with $\lambda =\lambda_1-\beta E$ admits a unique real-valued solution $\phi \in H^1(\R)$ close to $\sum _{j\in \Z} F_j u_j$, i.e., 
\be
\left \| \phi -\sum_{j\in \Z} F_j u_j \right \|_{H^1}  \le C_{\hbar}, \label {Equation33b}
\ee
where the constant $C_{\hbar} = \asy \left ( e^{-\alpha/\hbar } \right ) $, for some fixed $\alpha \in (0, S_0/2)$,  is exponentially small when $\hbar \to 0$.
\end{theorem}

\begin {remark}
The solution $\phi$ given in Theorem \ref {Theorem0} is not normalized to $1$. \ However, recalling that $\Vector {F}$ is normalized to $1$ then it follows that $\| \phi \|_{L^2} = 1 + C_\hbar$, where $C_\hbar = \asy (e^{-\alpha/\hbar})$, for some $0 < \alpha < S_0$, is exponentially small as $\hbar$ goes to zero. \ Hence, Theorem \ref {Theorem0} yields to a normalized solution provided that we replace $\gamma$ by $\gamma (1+C_\hbar)^{2\sigma}$.
\end {remark}

\begin {remark}
A similar result has been previously obtained \cite {P} in the case of $\sigma =1$. \ In particular, in \cite {P} the estimate of the remainder terms (\ref {Equation33b}) and (\ref {Equation39}) are given in the energy norm defined as (assuming that $V\ge 0$)
\bee
\| \psi \|^2_{{\mathcal H}^1} := \| \psi \|^2_{{H}^1} + \hbar^{-2} \langle \psi , V \psi \rangle \, . 
\eee
In order to get such estimates an assumption on the periodic potential $V$ is required; in fact, the results by \cite {P} do not apply for any periodic potential, but only for periodic potentials  with degenerate minima with infinite order (i.e. $V^{(n)} (x_j) =0$ for any $n \in \N$); that is practically piecewise constant periodic potentials. \ With more details, \cite {P} considers the NLSE in the form $i u_t =-u_{xx} + \hbar^{-2} Vu + |u|^2 u$ where the stationary solutions of the DNLS give, by means of the Wannier functions, the solutions of the NLSE. \ In order to get this result several assumptions on the band functions are given, in particular it is required that the mean value of the first band function of the Bloch operator $-\partial_{xx} + \hbar^{-2} V$ is bounded  when $\hbar \to 0$. \ In fact, WKB arguments imply that this condition is fulfilled only when $V$ has a minimum point infinitely many degenerate, if not the mean value of the first band goes as $\hbar^{-\alpha}$ for some $\alpha >0$ and then the assumption above is not satisfied. \ In \cite {PSc} the extension to smooth periodic potentials is given (again in the case of cubic nonlinearity), provided that the band functions satisfy to a technical assumption concerning the cubic terms (see Assumption 2-(iii) \cite {PSc}).
\end {remark} 

As a result of Theorem \ref {Theorem0}, Lemma \ref {lemma11} (below) and Remark \ref {Remark7Ter} then the phase transition from delocalized states, for small $|\eta |$ (superfluidity phase), to localized states, for large $|\eta|$ (Mott insulator phase) follows:

\begin{corollary} \label {Cor1} 
Let Hypotheses 1 and 2 be satisfied, and let $\sigma \ge 1/2$. \ If the absolute value of the effective nonlinearity parameter $\eta$ is large enough then equation (\ref {Equation31}) has a family of stationary solutions, where each solution is localized on one single well, with energy $\lambda = \lambda_1 - \beta E = \lambda_1 -\beta (2 - \eta + o(1))$ in the limit of large $|\eta |$. 
\end{corollary}

\subsection {Proof of Theorem \ref {Theorem0}}

By Remark \ref {Remark7Ter}, a solution of (\ref{Equation31})-if exists-is anyway real-valued up to gauge choices. Thus, in order to prove Theorem \ref {Theorem0}, we look for a real-valued solution $\phi (x)$ of the stationary equation (\ref {Equation31}) under the following form
\be
\phi = \phi^1 + \phi^\perp \, , 
\label {Equation34}
\ee
where $\phi^1 = \Pi \phi$ is the projection of the first band and where $\phi^\perp = \Pi^\perp \phi $, $\Pi^\perp = \I - \Pi$, is the projection of $\phi$ on the other bands. \ By construction, it turns out that 
\bee
\sigma (H \Pi ) =[\alpha_1 ,\beta_1] \ \mbox { and } \ \sigma (H \Pi^\perp ) \subseteq [\alpha_2 , +\infty ) 
\eee
where 
\be
C \hbar \le \alpha_2 - \beta_1 \le \frac {1}{C} \hbar \label  {Equation35}
\ee
for some positive constant $C$. \ By substituting (\ref {Equation34}) into (\ref {Equation31}) and projecting on the first band and on the other bands we obtain
\be
\left \{
\begin {array}{lcl} 
\lambda \phi^1 &=& H \phi^1 + \gamma \Pi |\phi|^{2\sigma} \phi \\ 
\lambda \phi^\perp &=& H \phi^\perp + \gamma \Pi^\perp |\phi|^{2\sigma} \phi
\end {array}
\right. \, .
\label {Equation36}
\ee
If we set 
\bee
\phi^1 (x )= \sum_{j \in \Z } c_j u_j (x)  \, , 
\eee
then the first equation of (\ref {Equation36}) reduces to the following system
\be
\lambda c_j = \langle u_j , H \phi^1 \rangle + \gamma L_j , \ L_j := \langle u_j , \Pi  |\phi|^{2\sigma } \phi \rangle \label {Equation37}
\ee
where 
\bee
\langle u_j , H \phi^1 \rangle = \sum_k c_k \langle u_j , H u_k \rangle \, . 
\eee
We underline that
\bee
\| \phi^1 \|_{L^2} = \| \Vector {c} \|_{\ell^2}
\eee
because the set of vectors $\{ u_j \}$ is an orthonormal basis of $\Pi L^2 (\R )$. \ The matrix with elements $\langle u_j , H u_k \rangle$ can be written as 
\bee
(\langle u_j , H u_k \rangle ) = \lambda_1 \I - \beta {\mathcal T} + \tilde D 
\eee
because of (\ref {Equation24}), Lemmata \ref {Lemma4} and \ref {Lemma5} and Remarks \ref {Remark3} and \ref {Remark4}, ${\mathcal T}$ is the tridiagonal Toeplitz matrix
\bee
({\mathcal T})_{j,k} = 
\left \{
\begin {array}{ll}
0 & \ { if }\ |j-k| \not= 1 \\
1 & \ { if }\ |j-k| = 1
\end {array}
\right. \, 
\eee
and the remainder term $\tilde D$, obtained collecting the elements $w_{j,k}$, when $|j-k|>1$, and $r_{j,k}$, is such that
\bee
\| \tilde D \|_{{\mathcal L} (\ell^p \to \ell^p )} \le C e^{-(S_0+\alpha ) /\hbar } \, , \ p \in [1,+\infty ] \, ,  
\eee
for some $0< \alpha <S_0$ and $C=C_p$.

First, we justify the existence of $\phi^{\perp}.$ 

\begin{theorem} \label{Theorem1} 
Assume $\sigma >0$ and fix any number $E \in \R$ and $\delta_0 >0$. \ For any $\Vector{c}=(c_j)_{j\in \Z} \in \ell^1(\Z) $ with $\|\Vector{c}\|_{\ell^1 (\Z)} \le \delta_0$, there exists a unique smooth map $\hat{\phi} : \ell^1(\Z) \times \R \to H^1(\R)$ such that $\phi^{\perp}=\hat{\phi}(\Vector{c},\hbar)$ is a solution of the second equation of (\ref{Equation36}) for small $\hbar>0$. Moreover, it is exponentially small as $\hbar \to 0$ in the sense that for any $0 < \nu < S_0$ there exists a positive constant $C>0$, independent of $\hbar$, such that 
\be
\|\phi^{\perp}\|_{H^1} \le C e^{-(S_0 - \nu )/\hbar} \|\Vector{c}\|_{\ell^1}^{2\sigma+1} \, .  \label {Equation39}
\ee
\end{theorem}

{\it Proof.} \ Note that the operator $H-(\lambda_1+\beta E)$ on $\Pi^{\perp}L^2$ has the inverse for $\hbar$ sufficiently small thanks to  (\ref{Equation35}). \ Precisely, by the functional calculus, since $V$ is a bounded potential then there exists a constant $C_1>0$ independent of $\hbar$ such that  
\bee
\|(H-(\lambda_1+\beta E))^{-1} \Pi^{\perp}\|_{\mathcal{L}(L^2 \to H^1)} \le C_1 \hbar^{-1}.
\eee
Then the second equation of (\ref{Equation36}) may be written as
\be
\phi^{\perp}=F(\phi^{\perp}), \label {Equation40}
\ee
where we set $\lambda_1 = \lambda + \beta E$ and 
\bee
F(\phi^{\perp}) =-\gamma (H-(\lambda_1+\beta E))^{-1} \Pi^{\perp}|\phi|^{2\sigma} \phi, \quad \phi=\phi^1 +\phi^{\perp}.  
\eee
We wish to make $F$ be a contraction mapping in a complete metric space. \ With this aim, we make here several remarks about the property of 
$\sum_{j \in \Z} c_j u_j (x)$. 

First, it follows from Lemma \ref{Lemma3}  that 
\bee
\|\psi_j - u_j \|_{L^p} \le C e^{{-(S_0-\nu)}/{\hbar}}
\eee
for $p \in [2,\infty]$. \ Furthermore, Lemma \ref {Lemma1} gives us the following estimate.  
\bee
 \| u_j \|_{L^p} \le C \hbar^{-\frac{p-2}{4p}}+ C e^{{-(S_0-\nu)}/{\hbar}}.
\eee
Remind that $C$ does not depend on $\hbar$ nor on the index $j$. \ Then, for any $p \in [2,\infty]$ 
\bee
\left \|\sum_{j \in \Z} c_j u_j \right \|_{L^p} &\le& \sum_{j\in \Z} |c_j| \|u_j\|_{L^p} \\  
&\le & \sum_{j \in \Z} |c_j| \Big\{ C \hbar^{-\frac{p-2}{4p}}+ C e^{{-(S_0-\nu)}/{\hbar}}\Big\} \\
&\le & \|\Vector{c}\|_{\ell^1} \Big\{ C \hbar^{-\frac{p-2}{4p}}+ C e^{{-(S_0-\nu)}/{\hbar}}\Big\} \\ 
&\le& \delta_0 \Big\{ C \hbar^{-\frac{p-2}{4p}}+ C e^{{-(S_0-\nu)}/{\hbar}}\Big\}.
\eee

We set 
\bee
\delta(\hbar)=\Big(\frac{\hbar}{4 \gamma 2^{2\sigma+1} C_1}\Big)^{\frac{1}{2\sigma}}.
\eee
Fix $\hbar>0$ so small that (we recall that $\gamma$ is exponentially small because of Hyp. \ref {Hyp2})
\bee
\frac{\gamma 2^{2\sigma+1} C_1}{\hbar}< 1/4, \quad 
\left \|\sum_{j \in \Z} c_j u_j \right \|_{L^{2(2\sigma+1)}}^{2\sigma+1} < \delta(\hbar), \quad 
\frac{\gamma 2^{2\sigma} C_1}{\hbar} 
\|\sum_{j \in \Z} c_j u_j\|_{L^{4\sigma}}^{2\sigma}<1/4. 
\eee

We shall show that $F$ is a contraction map in 
\bee
K= \{\xi^{\perp} \in H^1(\R) \cap \Pi^{\perp}L^2(\R) : \|\xi^{\perp}\|_{H^1} \le \delta(\hbar)\}.
\eee
Indeed, for any $\xi=\xi^1 +\xi^{\perp}$, $\eta=\eta^1 +\eta^{\perp}$,
\bee
\| F(\xi^{\perp}) \|_{H^1} 
&\le& \frac{\gamma C_1}{\hbar}  \||\xi|^{2\sigma} \xi\|_{L^2} \\
&\le& \frac{\gamma 2^{2\sigma+1} C_1}{\hbar} (\|\xi^1 \|_{L^{2(2\sigma+1)}}^{2\sigma+1} +\|\xi^{\perp}\|_{L^{2(2\sigma+1)}}^{2\sigma+1})\\  
&\le& \frac{\gamma 2^{2\sigma+1} C_1}{\hbar} \delta(\hbar) + \frac{1}{4} \delta(\hbar) < \delta(\hbar), \\ 
\|F(\xi^{\perp})-F(\eta^{\perp})\|_{H^1} 
&\le& \frac{\gamma C_1 2^{2\sigma}}{\hbar}  (\|\xi\|_{L^{4\sigma}}^{2\sigma} + \|\eta\|_{L^{4\sigma}}^{2\sigma}) \|\xi^{\perp} - \eta^{\perp} \|_{H^1} \\
&<& \Big(2\cdot \frac{1}{4} + \frac{\gamma C_1 2^{2\sigma}}{\hbar} \frac{\hbar}{4 \gamma 2^{2\sigma+1} C_1} \Big)\|\xi^{\perp} - \eta^{\perp} \|_{H^1} \\
&<& \frac{3}{16} \|\xi^{\perp} - \eta^{\perp} \|_{H^1}    
\eee
Then there exists a unique solution $\hat{\phi} = \hat {\phi } (\Vector {c} , \hbar ) \in K$ of Eq. (\ref {Equation40}) for small $\hbar>0$. \ Moreover, by the construction of the solution (see \cite{FS}), 
\bee
\|\xi^{\perp}\|_{H^1} \le \frac{1}{1-3/16} \frac{\gamma C_1}{\hbar} \||\xi^1|^{2\sigma}\xi^1\|_{H^1} \le  C e^{-(S_0 - \nu)/\hbar} \|\Vector{c}\|_{\ell^1}^{2\sigma+1}
\eee
for any $\nu \in (0, S_0)$ because of Remark \ref {Remark7}. \ Second inequality holds true because 
\bee
\||\xi^1|^{2\sigma} \xi^1\|_{H^1}\le \|\xi^1\|_{L^{4\sigma+2}}^{2\sigma +1} + \|\nabla (|\xi^1|^{2\sigma} \xi^1) \|_{L^2} \, ; 
\eee
indeed, we may estimate the second term as follows.
\bee
\|\nabla (|\xi^1|^{2\sigma} \xi^1) \|_{L^2} 
& \le & C_{\sigma} \| \nabla \xi^1 \| \cdot  \|\xi^1 \|^{2\sigma }_{L^\infty} \\ 
&\le &  C'_{\sigma} \hbar^{-1/2} \| \Vector {c}\|_{\ell^1} \cdot  \|\xi^1 \|^{2\sigma }_{L^\infty}
\eee
since
\bee
\| \nabla \xi^1 \|_{L^2} = \left \| \sum_{j} c_j \nabla u_j \right \|_{L^2} \le \sum_{j} | c_j | \| \nabla u_j \|_{L^2} \le C \hbar^{-1/2} \| \Vector {c} \|_{\ell^1} \,  
\eee
where we have used (\ref{Equation25bis}) and (\ref{Equation6}) for the second inequality, and the estimates
\bee
\|\xi^1 \|_{L^{\infty}}^{2\sigma} 
= \|\sum_{j} c_j u_j \|_{L^{\infty}}^{2\sigma} 
\le \Big( \sum_{j} |c_j| \|u_j\|_{L^{\infty}} \Big)^{2\sigma} 
\le C \hbar^{-\sigma/2} \|\Vector{c}\|_{\ell^1}^{2\sigma} \, , 
\eee
for some positive constant $C$ depending on $\sigma$, complete the proof.
\hfill\qed
\vspace{3mm}

The next lemma is concerned with the first equation of (\ref{Equation36}). \ The term $L_j$ in (\ref{Equation37}) may be expressed as follows 
\be
L_j= \|u_j\|_{L^{2\sigma+2}}^{2\sigma+2}|c_j|^{2\sigma} c_j + f_j, \quad  
f_j = f_j (\Vector {c} , \phi^\perp )\, . \label {formula1}
\ee
We set $\Vector {f} (\Vector {c}, \phi^\perp ) =\left ( f_j (\Vector {c}, \phi^\perp ) \right )_{j\in \Z}$, where $\phi^\perp = \hat \phi (\Vector {c}, \hbar )$ is given in Theorem \ref {Theorem1}.  

\begin {lemma} \label{lemma6}
Let $\hbar$ sufficiently small such that Theorem \ref{Theorem1} holds. \  Assume $\sigma \ge 1/2$, and  fix any number  $\delta_0 >0$. \ For any $\Vector{c}=(c_j)_{j\in \Z} \in \ell^1(\Z) $ with $\|\Vector{c}\|_{\ell^1 (\Z)} \le \delta_0$, we have  
\be
\|\Vector {f} (\Vector {c} , \phi^\perp )\|_{\ell^1}  
\le C \max\{\delta_0^{(2\sigma +1)^2}, \delta_0^{4\sigma+1} \} e^{- (S_0-\nu ) /\hbar} \, ,  \label {formula2}
\ee
for any $\nu \in (0, S_0)$. \ The positive constant $C$ is independent of $\hbar$. 
\end {lemma}

{\it Proof of Lemma \ref {lemma6}.} \ As a first step we remark that $\phi^1$ satisfies the following estimates
\bee
\| \phi^1 \|_{L^2} = \| \Vector {c} \|_{\ell^2} \le \delta_0 
\eee
and
\be 
\| \nabla \phi^1 \|_{L^2} \le C\delta_0 \hbar^{-1/2} \, . \label {formula2Bis}
\ee
The first inequality simply comes from the inequality $\| \Vector {c} \|_{\ell^2} \le  \| \Vector {c} \|_{\ell^1} \le \delta_0$. \ Concerning inequality (\ref {formula2Bis}) we simply remark that
\bee
\| \nabla \phi^1 \|_{L^2} = \left \| \sum_j c_j \nabla u_j \right \|_{L^2} \le \sum_j |c_j|\, \| \nabla u_j \|_{L^2} \le C \hbar^{-1/2} \| \Vector {c} \|_{\ell^1} 
\eee
Hence,
\be
\| \phi^1 \|_{L^\infty} \le C \| \phi^1 \|_{L^2}^{1/2}  \| \nabla \phi^1 \|_{L^2}^{1/2}  \le C \hbar^{-1/4} \| \Vector {c} \|_{\ell^1} \, . \label {F1}
\ee

Now, in order to prove the Lemma we set, for $j\in \Z$, 
\bee
f_j(\Vector{c},\phi^{\perp})=f_{j,1}(\Vector{c}) + f_{j,2}(\Vector{c}, \phi^{\perp})
\eee
where
\bee
f_{j,1}(\Vector{c}) = \langle u_j , |\phi^1|^{2\sigma } \phi^1 \rangle - \|u_j\|_{L^{2\sigma+2}}^{2\sigma+2} |c_j|^{2\sigma} c_j 
\eee
and
\be
 f_{j,2} (\Vector {c}, \phi^\perp) = \langle u_j , |\phi|^{2\sigma } \phi - |\phi^1|^{2\sigma }  \phi^1 \rangle. \label {Equation49Bis}
\ee
In order to estimate the $\ell^1$-norm of $(f_{j,2}(\Vector{c},\phi^{\perp}))_{j\in \Z}$ we remark that
\bee
\sum_j |f_{j,2}(\Vector{c},\phi^{\perp})| &\le &  \left \langle \sum_j |u_j| , \left | |\phi|^{2\sigma } \phi - |\phi^1|^{2\sigma }  \phi^1 \right | \right \rangle \\ 
&\le & \left \| \sum_j |u_j| \right \|_{L^\infty} \|  |\phi|^{2\sigma } \phi - |\phi^1|^{2\sigma }  \phi^1 \|_{L^1}
\eee
where the first term is estimated by Lemma \ref {lemma10}-ii. and where, for what concerns the r.h.s., we remark that 
\be
|\phi|^{2\sigma } \phi - |\phi^1|^{2\sigma }  \phi^1 =  |\phi^1 + \phi^\perp |^{2\sigma }  \phi^\perp +  \left [ |\phi^1 + \phi^\perp |^{2\sigma } - |\phi^1|^{2\sigma } \right ] \phi^1. \label {formula3}
\ee
We make use of inequalities (\ref{Equation39}) and (\ref {F1}) obtaining that 
\be
&& \left \| |\phi|^{2\sigma } \phi - |\phi^1|^{2\sigma }  \phi^1 \right \|_{L^1} \le \\
&& \le C \left [ \| \phi^1\|_{L^\infty}^{2\sigma-1} \| \phi^1 \|_{L^2}+ \| \phi^\perp \|_{L^\infty}^{2\sigma-1} \| \phi^\perp \|_{L^2} \right ] \| \phi^\perp \|_{L^2}  \nonumber \\ 
&& \ \ + C \| \phi^1\|_{L^2} \left [ \| \phi^1\|_{L^\infty}^{2\sigma-1} + \| \phi^\perp \|_{L^\infty}^{2\sigma-1} \right ] \| \phi^\perp \|_{L^2} \nonumber \\
&& \le  C \left [ \hbar^{(2\sigma -1)/4} \| \Vector {c} \|_{\ell^1}^{2\sigma} + e^{- 2\sigma (S_0 - \nu)/\hbar} \| \Vector {c} \|_{\ell^1}^{2\sigma (2\sigma +1)} \right ] e^{- (S_0 - \nu)/\hbar} \| \Vector {c} \|_{\ell^1}^{(2\sigma +1)}  \nonumber \\ 
&& \ \ + C \left [ \hbar^{(2\sigma -1)/4} \| \Vector {c} \|_{\ell^1}^{2\sigma} + e^{- 2\sigma (S_0 - \nu)/\hbar} \| \Vector {c} \|_{\ell^1}^{(2\sigma)^2} \right ] e^{- (S_0 - \nu)/\hbar} \| \Vector {c} \|_{\ell^1}^{(2\sigma +1)} \nonumber \\
&& \le  C \max [ \| \Vector {c} \|_{\ell^1}^{2\sigma(2\sigma+1)} ,\| \Vector {c} \|_{\ell^1}^{2\sigma} ] \| \Vector {c} \|_{\ell^1}^{(2\sigma+1)} e^{- (S_0 - \nu)/\hbar} \nonumber \\ 
&& \le C \max\{\delta_0^{(2\sigma +1)^2}, \delta_0^{4\sigma+1}\} e^{- (S_0 - \nu)/\hbar} \, . \nonumber
\ee
Hence,
\bee
\| f_{j,2} (\Vector {c}, \phi^\perp) \|_{\ell^1} = \sum_j \left | \langle u_j , |\phi|^{2\sigma } \phi - |\phi^1|^{2\sigma }  \phi^1 \rangle \right | \le C e^{-(S_0 - \nu )/\hbar } \max\{\delta_0^{(2\sigma +1)^2}, \delta_0^{4\sigma+1}\} 
\eee 
for any $\nu >0$ and for some $C>0$. 

In order to estimate the term $(f_{j,1}(\Vector{c}))_{j\in \Z}$, which does not contain the vector $\phi^\perp$, let us denote, for a fixed $j \in \Z$, 
\bee
\tilde \phi^1 = \phi^1 - c_j u _j = \sum_{m\not= j} c_m u_m .
\eee
We see that $f_{j,1}(\Vector{c})$ is given by a finite sum of terms like $c_j \langle u_j , \tilde \phi^1 |\phi^1 |^{2\sigma} \rangle $ and $c_j \langle u_j ,\overline { \tilde \phi^1} (\phi^1)^{2\sigma+1}( \bar \phi^1 )^{2(\sigma-1)} \rangle $; therefore we only have to estimate such a kind of term. \ Indeed, by Lemma \ref {lemma10}-i.
\bee
 |\langle u_j , \tilde \phi^1 |\phi^1 |^{2\sigma} \rangle |  
&\le &  \sum_{m \not= j} |c_m| \| u_j u_m \|_{L^1} \| \phi^1 \|_{L^\infty}^{2\sigma}  \le \left [  \sum_{m \not= j} |c_m| \| u_j u_m \|_{L^1} \right ] C \hbar^{-\sigma} \\ 
& \le & \sum_{m\not= j } |c_m |  C e^{- \left [ |j-m| (S_0-\nu )- \tilde \nu \right ]/\hbar } 
\eee
for some $\tilde \nu = \asy (\nu )$. \ Then it follows that
\bee
\sum_j  |\langle u_j , \tilde \phi^1 |\phi^1 |^{2\sigma} \rangle |  
&\le & C \sum_j \sum_{m\not= j} |c_m| e^{-[|j-m| (S_0 - \nu ) - \tilde \nu ] /\hbar } \\ 
&\le & C \sum_m |c_m| \sum_{j\not= m} e^{-[|j-m| (S_0 - \nu ) - \tilde \nu ] /\hbar } \\ 
&\le & C \sum_m |c_m|  e^{-[ (S_0 - \nu ) - \tilde \nu ] /\hbar } \le C \| \Vector {c} \|_{\ell^1}  e^{-[ (S_0 - \nu ) - \tilde \nu ]  /\hbar }\\
&\le & C \| \Vector {c} \|_{\ell^1}  e^{- (S_0 - \nu )   /\hbar } \le C \delta_0  e^{- (S_0 - \nu )   /\hbar } 
\eee
where we set $\nu + \tilde \nu \to \nu$ and $\| \Vector {c} \|_{\ell^1} \le \delta_0$. \ The Lemma is so proved.
\hfill\qed 
\vspace{3mm}

Note that,  
\bee
\|u_j\|_{2\sigma+2}^{2\sigma+2}=\|T^{(j)} u_0 \|_{2\sigma+2}^{2\sigma+2}=\|u_0 \|_{2\sigma+2}^{2\sigma+2}= C_0 \,  
\eee
where $T$ is the translation operator defined in Remark \ref{NewRemark}. \ Thus, the system (\ref {Equation36}) takes the form
\be
\left \{
\begin {array}{lcl}
\lambda {c_j} &=& \lambda_1 {c}_j - \beta \left ( {\mathcal T}\Vector{c} \right )_j + ( \tilde D \Vector {c} )_j + \gamma C_0 |{c}_j|^{2\sigma } {c}_j +   \gamma f_{j} (\Vector{c},\phi^\perp) \\
\lambda \phi^\perp &=& H \phi^\perp + \gamma \Pi^\perp |\phi|^{2\sigma} \phi
\end {array}
\right. \, . 
\label {Equation38}
\ee
Dividing by $\beta$, the first equation of (\ref{Equation38}) may be re-written under the following form; 
\be
E {c_j} =  -  \left ( {\mathcal T}\Vector{c} \right )_j + \frac {1}{\beta} ( \tilde D \Vector {c} )_j +  \eta |{c}_j|^{2\sigma} {c}_j +\frac{\gamma}{\beta} f_{j}(\Vector{c},\phi^\perp)  \label {H3}
\ee
where the remainder term $\frac{\gamma}{\beta} f_{j}(\Vector{c},\phi^\perp)$ is estimated in Lemma \ref{lemma6}, and where $\tilde D$ is estimated in Remark \ref {Remark4}. \ The approximated system of equations obtained by neglecting the remainder terms $\frac {1}{\beta} ( \tilde D \Vector {c} )_j$ and 
$\frac{\gamma}{\beta} f_{j}(\Vector{c},\phi^\perp)$ in the Eq. (\ref {H3}) thus takes the form (\ref {Equation32}) and it is a discrete nonlinear 
Schr\"odinger equation equivalent to the so called \emph {Bose-Hubbard} model.

We are now in position to complete the proof of Theorem \ref{Theorem0}. 
\vspace{3mm}

{\it Proof of Theorem \ref{Theorem0}}. \ Recall that we look for a real-valued solution $\phi $ of the stationary equation (\ref {Equation31}). \ From (\ref {H3}) we consider the following mapping 
\bee
\mathcal{F} : \ell^1_{\R} \times \R \to \ell^1_{\R}, \quad (\Vector{c},\gamma) \mapsto  
\mathcal{F}(\Vector{c}, \gamma)
\eee
such that  
\bee
\mathcal{F}(\Vector{c}, \gamma)=(\mathcal{F}_j(\Vector{c}, \gamma))_{j\in \Z}
\eee
and
\bee
\mathcal{F}_j(\Vector{c}, \gamma)=  E {c_j} +  \left ( {\mathcal T}\Vector{c} \right )_j - \frac {1}{\beta} \left ( \tilde D \Vector {c} \right )_j - \eta |{c}_j|^{2\sigma} {c}_j - \frac{\gamma}{\beta} f_{j} (\Vector{c},\phi^\perp)  \, ,
\eee
where $\phi^\perp = \hat \phi (\Vector {c},\hbar)$ is the solution of equation (41) given in Theorem \ref {Theorem1}. \ This map is well defined; indeed 
we have already seen in Lemma \ref{lem:realvalued} that $u_j$ can be chosen to be real-valued, thus for $\Vector{c} \in \ell^1_{\R}$,  
$f_j (\Vector{c}, \phi^{\perp})$ takes real values by construction. \ This map is indeed $C^1$ in $(\Vector{c},\gamma)$ for any $\sigma>0$; here, we show that the map $\Vector {f} : \ell^1_\R \times H^1 \to \ell^1_\R$ is $C^1$ in $\Vector{c}$. \ In order to prove it, we consider, at first, 
the map $\Vector {c}\in \ell^1_\R \mapsto \Vector {f}_1 (\Vector {c}) = \left ( f_{1,j} (\Vector {c}) \right )_j$ where $f_{1,j}$ is defined in the proof of Lemma \ref {lemma6} as 
\bee
f_{1,j}(\Vector{c}) = \langle u_j , |\phi^1|^{2\sigma} \phi^1 \rangle - C_0 |c_j|^{2\sigma} c_j \, . 
\eee
For any $\Vector {c}$, $\Vector {h} \in \ell^1_\R$ we define the linear map from $\ell^1_\R$ to $\ell^1_\R$ 
\bee
A_1 (\Vector {h}) = \left ( (2\sigma +1) \left ( \sum_k \langle u_j , |\phi^1|^{2\sigma} u_k \rangle h_k - C_0 |c_j|^{2\sigma} h_j \right ) \right )_{j\in \Z} \, . 
\eee
By computation we directly have that $D_{\Vector{c}} \Vector {f}_1 (\Vector {c}) = A_1$. \ For what concern the boundedness of $A_1$ we have that the same arguments given in the proof of Lemma \ref {lemma6} and the asymptotic estimate of $\| u_j u_k \|_{L^1}$ give that 
\be
\| A_1\|_{{\mathcal L}(\ell^1 \to \ell^1)} \le C_{\sigma} (1+\|\Vector{c}\|_{\ell^1}^2 +\|\Vector{c}\|_{\ell^1}^{2\sigma}) e^{-(S_0 - \nu )/\hbar }. \nonumber
\ee 
In particular $D_{\Vector{c}} \Vector {f}_1 (\Vector {c})$ is continuous in $\Vector{c}$. \ Concerning $\Vector {f}_2 (\Vector {c}, \phi^\perp )
=(f_{j,2}(\Vector{c}, \phi^{\perp}))_{j\in \Z}$, which is defined by (\ref {Equation49Bis}), the same argument applies, and where the map $\Vector {c} \mapsto \phi^\perp = \hat \phi (\Vector {c} , \hbar )$ is smooth by means of the implicit function theorem we have used in the proof of Theorem \ref {Theorem1}. \ Therefore we have 
\be 
\|D_{\Vector{c}} \Vector{f}(\Vector{c}, \hat{\phi})\|_{{\mathcal L}(\ell^1 \to \ell^1)}
\le C_{\sigma} (1+\|\Vector{c}\|_{\ell^1}^2 +\|\Vector{c}\|_{\ell^1}^{2\sigma}) e^{-(S_0-\nu)/\hbar}. \label {FormulaCircle}
\ee
From Hyp.2 it turns out that $\frac {\gamma }{\beta } = \frac {\eta }{C_0}$, where $C_0 = \| u_0 \|_{L^{2\sigma +2}}^{2\sigma +2} = \asy (\hbar^{-\sigma /2} )$ and $\eta$ goes to a real value in the multi-scale limit; from these facts and from Lemma \ref {lemma6} and Remark \ref{Remark4} we have that 
\be
\left \| \frac{\gamma}{\beta} \Vector {f} (\Vector{c},\phi^\perp) \right \|_{\ell^1} = O\left ( e^{-(S_0-\nu)/\hbar} \right ) \label {F71}
\ee
and 
\be
\frac {1}{\beta} \| \tilde D \Vector {c} \|_{\ell^1} \le \frac {C}{\beta} e^{-(S_0 + \alpha )/\hbar } \| \Vector {c} \|_{\ell^1} 
\ee
for some $0< \alpha <S_0$ and $C>0$, any $0 < \nu <S_0$ and any $\Vector {c} \in \ell^1_\R (\Z)$ such that $\| \Vector {c} \|_{\ell^1} \le \delta_0$ fixed. \ Hence, 
\bee
 \sup_{\| \Vector {c} \|_{\ell^1} \le \delta_0} 
\left \| \frac{\gamma}{\beta} \Vector {f}(\Vector{c},\hat \phi (\Vector {c} , \hbar ) ) + \frac {1}{\beta} (\tilde D \Vector {c} ) \right \|_{\ell^1} =  \asy (e^{-\alpha /2\hbar } )  =: C_\hbar \, . 
\eee
Now, we fix $\delta_0 \ge 1$ and $\mu_{\hbar} :=  e^{-\alpha /2\hbar } $, and we define now the regular mapping
\bee
\mathcal{G} : \ell^1_{\R} \times \R \to \ell^1_{\R}, \quad (\Vector{c},y) \mapsto \mathcal{G}(\Vector{c},y)  
\eee
such that  
\bee
\mathcal{G}(\Vector{c},y)=\{\mathcal{G}_j(\Vector{c},y)\}_{j\in\Z}
\eee
and 
\bee
\mathcal{G}_j(\Vector{c}, y)=  E {c_j} +  \left ( {\mathcal T}\Vector{c} \right )_j - \eta  |{c}_j|^{2\sigma} {c}_j + y g_j (\Vector {c} , \hbar ) \, 
\eee
where 
\bee
g_j (\Vector{c},\hbar) := -  \frac{1}{\mu_{\hbar}} \left [\frac {\gamma}{\beta} {f}_j (\Vector{c},\hat{\phi}(\Vector{c},\hbar)) + \frac {1}{\beta} (\tilde D \Vector {c})_j  \right ]  
\eee
satisfies
\bee
\ \sup_{\| \Vector {c} \|_{\ell^1} \le \delta_0} \| \Vector{g} \|_{\ell_1} = \asy (1) \, , \ \mbox { as } \hbar \to 0 \, , \quad 
\Vector{g}(\Vector{c},\hbar)=\{g_j(\Vector{c},\hbar)\}_{j\in\Z} \, . 
\eee
By construction at $y=\mu_{\hbar}$ then $\mathcal{G}_j(\Vector{c}, y)$ coincides with $\mathcal{F}_j(\Vector{c}, \gamma )$. \ The map ${\mathcal G}$ satisfies $\mathcal{G}(\Vector{F}, 0)=0$, where $\Vector{F}=\{F_j \}_{j \in \Z} \in \ell^1_{\R} (\Z )$ is a non-trivial real-valued solution of (\ref {Equation32}), $\mathcal{G}(\Vector{c}, y)$ is $C^1$ in a neighborhood of $(\Vector{F},0)$, with the same reason from (\ref {FormulaCircle}), for any $\sigma>0$ and the linearized map 
\bee
D_{\Vector{c}} \mathcal{G}(\Vector{F},0) :\ell^1_{\R} \to \ell^1_{\R}
\eee 
can be written as 
\bee
D_{\Vector{c}} \mathcal{G}(\Vector{F},0) (\Vector {v} ) =L_+ \Vector {v}
\eee
and it is one-to-one and onto (see Hyp. 3). \ In fact the neighbourhood of the point $(\Vector{F},0)$ in which $C^1$ property of $\mathcal{G}(\Vector{c}, y)$ is held is independent of $\hbar$; let $\Vector{c}=\Vector{F}+\Vector{a}$ with $\|\Vector{a}\|_{\ell^1} \le r$, and $|y| \le \delta.$ Then, for any fixed small $\hbar>0$ and any $\epsilon>0$,  
\begin{eqnarray*}
&& \|D_{\Vector{c}} \mathcal{G} (\Vector{F}+\Vector{a}, y)\Vector{h}
-D_{\Vector{c}}\mathcal{G} (\Vector{F},0)\Vector{h}\|_{\ell^1} \\
&&\le |\eta| \|\{(|F_j+a_j|^{2\sigma}-|F_j|^{2\sigma})h_j\}_{j}\|_{\ell^1}
+|y|\frac{1}{\mu_{\hbar}} \|\{\frac{\gamma}{\beta} D_{\Vector{c}} f_j (\Vector{F}+\Vector{a})h_j
+\frac{1}{\beta} (\tilde{D} h)_j \}_j \|_{\ell^1} \\
&& \le  C' |\eta| r \|\Vector{h}\|_{\ell^1} 
+|y| \frac{1}{\mu_{\hbar}} C(1+r^2+r^{2\sigma}) e^{-\alpha/2\hbar} \|\Vector{h}\|_{\ell^1}. 
\end{eqnarray*}
Here we can take $r={C'}^{-1}{\epsilon}/{2|\eta|}$, $\delta=C^{-1} \epsilon /6$ (independent of $\hbar$).

Therefore, by the Implicit Function Theorem, there exist an $\hbar$-independent $\delta >0$ such that if $|y | \le \delta $ then there exists a  unique solution $\Vector {c}(y)$ in a $\ell^1$-neighborhood of $\Vector {F}$ satisfying $\mathcal{G} (\Vector{c}, y)=0$. \ Since $\mu_{\hbar}$ is of order $C_\hbar$ as $\hbar$ goes to zero, $y=\mu_{\hbar}$ is in the neighborhood for sufficiently small $\hbar$. \ Remind that the solution of $\mathcal{G} (\Vector{c},y)=0$ at $y=\mu_{\hbar}$ coincides with the solution of $\mathcal{F} (\Vector{c}, \gamma )=0$. \ Then we can conclude that there exists $\hbar^\star >0$ such that for any $\hbar < \hbar^\star$ there is a unique solution $\Vector{c} \in B_{C_{\hbar}}(\Vector{F}, \ell^1_\R )$ satisfying $\mathcal{F} (\Vector{c}, \gamma)=0$, for any $\gamma$ in a neighborhood of zero. \ Indeed it follows that the radius of the ball (neighborhood) of $\Vector{F}$ is of order 
$C_{\hbar}$ since the map $y \to \Vector {c}(y)$ is $C^1$. 

Then the Theorem follows since
\bee
\| \phi - \sum_j F_j u_j \|_{H^1} &\le & \| \phi^\perp \|_{H^1} + \left \| \phi^1 - \sum_j F_j u_j \right \|_{H^1} \\ 
&\le &  \| \phi^\perp \|_{H^1} + \left \|  \sum_j (c_j -F_j) u_j \right \|_{H^1} \\ 
&\le &  \| \phi^\perp \|_{H^1} + \left \| \Vector {c} - \Vector {F} \right \|_{\ell^1} \| u_0 \|_{H^1} \\ 
& \le & C e^{-(S_0 -\nu )/\hbar } + C e^{-\alpha /2\hbar} \le C e^{-\alpha /2\hbar} 
\eee
for any $0<\nu <S_0$ and for some $0 < \alpha < S_0$, which is exponentially small. \ Theorem \ref {Theorem0} is so proved.  
\hfill\qed 
\vspace{3mm}

\subsection {On the validity of Hyp.3 and proof of Corollary 1}

We start with a result concerning Hyp.3

\begin {lemma} \label {lemma11}
Let $\sigma>0$. \ Then Hyp.3 is satisfied for $|\eta |$ large enough.
\end {lemma}

{\it Proof.} \ As a first step we prove that $L_+$ is one-to-one by proving that for any $\eta_*>1$ there exists $\eta <0$ with $|\eta|\ge \eta_*$ such that
\bee
\|L_+ \Vector{v}\|_{\ell^1} \ge \frac{1}{2}\|\Vector{v}\|_{\ell^1}, \quad \forall \Vector{v} \in \ell^1_\R. 
\eee
Indeed, let $\Vector {F}= \Vector {F} (\eta )$ be the real-valued ground state solution of equation (\ref {Equation32}). \ Let $\Vector{G}^{\infty}$ be a solution of the following equation
\be
-|G^{\infty}_j|^{2\sigma} G_j^{\infty}=-G_j^{\infty}, \label {F70}
\ee
where $\Vector {G}^\infty \in \ell^p_\R$, $\| \Vector {G}^\infty \|_{\ell^p} =1$ for any $p \in [1,+\infty ]$. \ In fact, we have a family of solutions  $\Vector {G}^\infty = \Vector {G}^{\infty , j_0}$ for any $j_0 \in \Z$ where $G_j^{\infty , j_0}=\pm \delta_{j}^{j_0}$. \ Hereafter we fix a given value for $j_0$ and we denote by $\Vector {G}^\infty$ the corresponding solution. \ Recall Remark \ref{Remark7Ter} and we have a solution $\Vector {F} = \Vector {F} (\eta)$ of equation (\ref {Equation32}) such that $\| \Vector{F}(\eta) - \Vector{G}^{\infty}\|_{\ell^2} \le C |\eta|^{-1}$ and $\frac{E-2}{|\eta|} \to 1$ as $|\eta | \to \infty$. \ Define the following operators 
\bee
(\tilde L_+ \Vector{v})_j &=& -\frac{1}{|\eta|} (v_{j+1}+v_{j-1}) + \Big(\frac{E}{|\eta|}-(2\sigma+1)|F_j|^{2\sigma}\Big)v_j \\ 
(L_{\infty} \Vector{v})_j &=& (1-(2\sigma+1)|G_j^{\infty}|^{2\sigma})v_j.  
\eee
Then, we see, 
\bee
\|L_{\infty} \Vector{v}\|_{\ell^1} = \sum_{j\in \Z}|v_j -(2\sigma+1)|G_j^{\infty}|^{2\sigma} v_{j}| = \sum_{j \ne j_0} |v_j| + |-2\sigma v_{j_0}| 
\ge \min\{1, |2\sigma|\}\|\Vector{v}\|_{\ell^1}.
\eee
On the other hand, 
\bee
\|(\tilde L_+ - L_{\infty}) \Vector{v}\|_{\ell^1} 
&\le& \sum_{j\in \Z} \frac{1}{|\eta|} (|v_{j+1}|+ |v_{j-1}|+2|v_j|) \\
&& +\Big|\frac{E-2}{|\eta|}-1 \Big| \sum_{j\in\Z} |v_j| +(2\sigma+1) \sum_{j\in \Z} 
||F_j|^{2\sigma} - |G_j^{\infty}|^{2\sigma}||v_j|
\eee
The first term on the right hand side is bounded by $\frac{4}{|\eta|} \|\Vector{v}\|_{\ell^1} $ which tends to zero as $|\eta|$ goes to $\infty$. \ The third term is estimated as follows: if $2\sigma \ge 1$, 
\bee
\sum_{j\in \Z}||F_j|^{2\sigma}-|G_j^{\infty}|^{2\sigma}||v_j| 
&\le& (\sum_{j\in \Z}||F_j|^{2\sigma}-|G_j^{\infty}|^{2\sigma}|^2)^{1/2} \|\Vector{v}\|_{\ell^2}\\
& \le & C \Big\{\sum_{j} (|F_j|^{2(2\sigma-1)} + |G_j^{\infty}|^{2(2\sigma-1)}) |F_j-G_j^{\infty}|^2 \Big\}^{1/2} \|\Vector{v}\|_{\ell^1}\\
&\le & C' (\|\Vector{F}\|_{\ell^{\infty}}^{2\sigma-1} +\|\Vector{G}^{\infty}\|_{\ell^{\infty}}^{2\sigma-1}) \|\Vector{F}-\Vector{G}^{\infty}\|_{\ell^2} \|\Vector{v}\|_{\ell^1}.
\eee
If $0<2\sigma<1$, 
\bee
\sum_{j\in \Z}||F_j|^{2\sigma}-|G_j^{\infty}|^{2\sigma}||v_j| 
\le \|\Vector{F}-\Vector{G}^{\infty}\|_{\ell^{\infty}}^{2\sigma} \|\Vector{v}\|_{\ell^1}
\le \|\Vector{F}-\Vector{G}^{\infty}\|_{\ell^{2}}^{2\sigma} \|\Vector{v}\|_{\ell^1}.
\eee
Therefore, for any $\eta$ with $|\eta|>1$
\bee
\|L_+ \Vector{v}\|_{\ell^1} &=& |\eta| \|\tilde L_+ \Vector{v}\|_{\ell^1} 
\ge \|\tilde L_+ \Vector{v}\|_{\ell^1} 
\ge \|L_{\infty} \Vector{v}\|_{\ell^1} - \|(\tilde L_+ - L_{\infty})\Vector{v}\|_{\ell^1} \\
&\ge& (\min\{1,|2\sigma|\}-a_{\eta}-b_{\eta})\|\Vector{v}\|_{\ell^1} 
\eee
for some $0< a_{\eta}, b_{\eta} \ll 1$ if $|\eta| \gg 1$. 

By means of the previous result we have that $L_+$ is one-to-one since $\mbox {Ker} (L_+) = \{ 0 \}$. \ In order to prove that $L^+$ is onto we should require the estimate on the adjoint. \ More precisely, let $L_+^\star : \ell^\infty_\R \to \ell^\infty_\R $ be the adjoint of the linear and bounded operator $L_+ : \ell^1_\R \to \ell^1_\R$. \ We prove that for any $\eta_*>1$ there exists $\eta$, with $|\eta|\ge \eta_*$, such that 
\be
\|L_+^\star \Vector{v}\|_{\ell^\infty} \ge \frac{1}{2}\|\Vector{v}\|_{\ell^\infty}, \quad \forall \Vector{v} \in \ell^\infty_\R. \label {Adj}
\ee
From such an estimate it follows that (see Theorem II.19 by \cite {Br}) $L_+$ is onto.

In order to prove (\ref {Adj}), first of all we remark that $L_+$ acts as 
\bee
\left ( L_+ \Vector {c} \right )_j = \sum_{k} \alpha_{j,k} c_k 
\eee
where $\alpha_{j,k}=0$ if $|j-k|>1$, 
\bee
\alpha_{j,j+1}=\alpha_{j,j-1} = -1 \ \mbox { and } \ \alpha_{j,j} = E+ \eta (2\sigma +1) |F_j|^{2\sigma} \, . 
\eee
Since $\alpha_{j,k} = \alpha_{k,j}$ are real valued then the adjoint $L_+^\star$ formally acts on $\ell^\infty_\R$ as 
\bee
\left ( L_+^\star \Vector {v} \right )_j = \sum_{k} \alpha_{j,k} v_k \, . 
\eee
We denote, as before, $\Vector{G}^{\infty}$ the solution of equation (\ref {F70}) and, similarly, we define the operators $\tilde L_+$ and  $L_\infty$ on $\ell^\infty_\R$. \ Then, we see, 
\bee
\|L_{\infty} \Vector{v}\|_{\ell^\infty} = \sup_{j\in \Z}|v_j -(2\sigma+1)|G_j^{\infty}|^{2\sigma} v_{j}| = \sup_{j \ne j_0} |v_j| + |-2\sigma v_{j_0}| 
\ge \min\{1,|2\sigma|\} \|\Vector{v}\|_{\ell^\infty}.
\eee
On the other hand, 
\bee
\|(\tilde L_+ - L_{\infty}) \Vector{v}\|_{\ell^\infty}  &\le& \sup_{j\in \Z} \frac{1}{|\eta|} (|v_{j+1}|+ |v_{j-1}|+2|v_j|) \\
&& +\Big|\frac{E-2}{|\eta|}-1 \Big| \sup_{j\in\Z} |v_j| +(2\sigma+1) \sup_{j\in \Z} ||F_j|^{2\sigma} - |G_j^{\infty}|^{2\sigma}||v_j|
\eee
The first term on the right hand side is bounded by $\frac{4}{|\eta|} \|\Vector{v}\|_{\ell^\infty} $ which tends to zero as $|\eta|$ goes to $\infty$. \ In order to estimate the third term we remark that: if $2\sigma \ge 1$ 
\bee
\left | |F_j |^{2\sigma} - |G_j^\infty |^{2\sigma} \right | &\le & C  |F_j - G_j^\infty | \left [ |F_j|^{2\sigma -1} + |G_j^\infty|^{2\sigma-1}  \right ] \\ 
&\le & C  \| \Vector {F} -\Vector {G}^\infty \|_{\ell^\infty} \left [ \| \Vector {F}\|_{\ell^\infty}^{2\sigma -1} + \| \Vector {G}^\infty \|_{\ell^\infty}^{2\sigma-1}  \right ],
\eee
hence,
\bee
 \sup_{j\in \Z}||F_j|^{2\sigma}-|G_j^{\infty}|^{2\sigma}||v_j| \le  \| \Vector {F} -\Vector {G}^\infty \|_{\ell^\infty} \left [ \| \Vector {F}\|_{\ell^\infty}^{2\sigma -1} + \| \Vector {G}^\infty \|_{\ell^\infty}^{2\sigma-1}  \right ] \| \Vector {v} \|_{\ell^\infty},
\eee
and if $0<2\sigma<1$, 
\bee
 \sup_{j\in \Z}||F_j|^{2\sigma}-|G_j^{\infty}|^{2\sigma}||v_j| 
\le  \| \Vector {F} -\Vector {G}^\infty \|_{\ell^\infty}^{2\sigma} \| \Vector {v} \|_{\ell^\infty},
\eee
where $\| \Vector {G}^\infty \|_{\ell^\infty} =1$, $\| \Vector {F} \|_{\ell^\infty} \to 1$  and $\| \Vector {F} - \Vector {G}^\infty \|_{\ell^\infty} \to 0$ as $|\eta |\to + \infty$. \ Therefore, (\ref {Adj}) follows. 
\hfill\qed 
\vspace{3mm}

Here, we are ready to prove the Corollary.

{\it Proof of Corollary 1.} \ In order to prove the Corollary we only have to check that the linearized map $L_+ :\ell^1_\R \to \ell^1_\R$ is, for $\eta$ large enough, one-to-one and onto. \ This fact is proved in Lemma \ref {lemma11}. \ Hence, the map $L_+$ is invertible and we may apply the same arguments of the proof of Theorem \ref {Theorem0} where, for $|\eta |$ large enough, the solution $\Vector {F}$ of equation (\ref {Equation32}) is close to the solution $\Vector {G}^\infty$ of equation (\ref {F70}), which is fully localized on a single lattice cell (see Remarks \ref {Remark7Bis} and \ref {Remark7Ter}). \ The Corollary is so proved.
\hfill\qed

\end {document}